\documentclass[aps, prl,  floatfix, superscriptaddress, longbibliography, twocolumn]{revtex4-1} 
\usepackage{graphicx}
\usepackage{epstopdf}
\usepackage{bm, amsmath, amssymb}
\usepackage{float}
\usepackage{placeins}
\usepackage{color}
\usepackage{siunitx}
\usepackage{setspace}
\usepackage{bigints}

\newcommand{\ket}[1]{\left|#1\right>}
\newcommand{\bra}[1]{\left<#1\right|}

\newcommand{\ba}{\textbf{q}}
\newcommand{\bb}{\textbf{c}}
\newcommand{\bH}{\textbf{H}}

\newcommand{\bq}{\textbf{q}}
\newcommand{\bc}{\textbf{c}}
\newcommand{\bff}{\textbf{f}}
\newcommand{\bM}{\textbf{M}}

\begin{document}

\title{Gated conditional displacement readout of superconducting qubits}


\author{S. Touzard}
\email{steven.touzard@yale.edu}
\affiliation{Department of Applied Physics and Physics, Yale University, New Haven, CT 06520, USA}
\author{A. Kou}
\affiliation{Department of Applied Physics and Physics, Yale University, New Haven, CT 06520, USA}
\author{N.E. Frattini}
\affiliation{Department of Applied Physics and Physics, Yale University, New Haven, CT 06520, USA}
\author{V.V. Sivak}
\affiliation{Department of Applied Physics and Physics, Yale University, New Haven, CT 06520, USA}
\author{S. Puri}
\affiliation{Yale Quantum Institute, Yale University, New Haven, CT 06520, USA}
\author{A. Grimm}
\affiliation{Department of Applied Physics and Physics, Yale University, New Haven, CT 06520, USA}
\author{L. Frunzio}
\affiliation{Department of Applied Physics and Physics, Yale University, New Haven, CT 06520, USA}
\author{S. Shankar}
\affiliation{Department of Applied Physics and Physics, Yale University, New Haven, CT 06520, USA}
\author{M.H. Devoret}
\email{michel.devoret@yale.edu}
\affiliation{Department of Applied Physics and Physics, Yale University, New Haven, CT 06520, USA}

\date{\today}

\begin{abstract}

We have realized a new interaction between superconducting qubits and a readout cavity that results in the displacement of a coherent state in the cavity, conditioned on the state of the qubit. This conditional state, when it reaches the cavity-following, phase-sensitive amplifier, matches its measured observable, namely the in-phase quadrature. In a setup where several qubits are coupled to the same readout resonator, we show it is possible to measure the state of a target qubit with minimal dephasing of the other qubits. Our results suggest novel directions for faster readout of superconducting qubits and implementations of bosonic quantum error-correcting codes.

\end{abstract}

\maketitle

Measuring the state of a qubit is a fundamental operation of quantum physics and a primitive for building a universal quantum computer \cite{DiVincenzo2000}. Over the years, non-destructive strategies to measure one given system at the scale of a single quantum of energy have been devised and tested, first with Rydberg atoms \cite{Haroche2006}. In circuit quantum electrodynamics (cQED), such quantum non-demolition (QND) readout schemes are currently based on a dispersive interaction: the phase of a coherent state of a microwave pulse is shifted depending on the state of the qubit \cite{Blais2004, Wallraff2004, Gambetta2008}. In the best cases, this phase is then indirectly measured using a phase-sensitive amplifier to record the quadrature along which the two phase-shifted coherent states are separated. In order to achieve faster high-fidelity measurement, this separation can be augmented by increasing the number of probing photons. Unfortunately, in practice, driving with more photons induces unwanted qubit transitions and does not improve significantly the overall fidelity of the readout process \cite{Picot2008, Boissonneault2009, Slichter2012,Sank2016}. 

To circumvent the flaws of the RF dispersive qubit readout, a new paradigm has been proposed \cite{Kerman2013, Billangeon2015, Didier2015}, which consists of two ideas. First, the Z component (energy operator) of the qubit needs to be directly coupled to the quadrature measured by a phase-sensitive amplifier, which does not in principle degrade the signal-to-noise ratio (SNR). This bare interaction has been referred to as ``longitudinal'' \cite{Kerman2013,Billangeon2015}. Such interaction is, in fact, similar to that associated with radiation pressure in optomechanics \cite{Aspelmeyer2014}. Second, the interaction needs to be modulated in time, at the frequency of the readout cavity mode \cite{Didier2015}. This modulation of the coupling creates a displacement of the cavity that is conditioned on the state of the qubit. Input squeezed light can further enhance the sensitivity \cite{Didier2015,Eddins2018}. Alternatively, the coupling can be modulated in a stroboscopic way to avoid the back-action of the microwave field \cite{Viola1999, Clerk2010,Suh2014,Eddins2018}. The bare longitudinal interaction has been realized experimentally with superconducting circuits \cite{Vion2002, Roy2017,Eichler2018} but, in absence of the frequency modulation, it has not yet led to a QND microwave readout.

In this letter, we report the realization of  such a conditional displacement readout using detuned parametric pumping of the Josephson Hamiltonian of a transmon \cite{Leghtas2015}. This latter technique is a practical alternative to the flux modulation that has been proposed theoretically \cite{Kerman2013, Billangeon2015, Didier2015}. As shown in \cite{supp}, the time-dependent qubit-cavity Hamiltonian, in the doubly rotating frame, is given by
\begin{equation}
\frac{\bH_{\text{eff}}}{\hbar} = -\frac{\alpha}{2}\ba^{\dag 2}\ba^2 + \zeta(t)(\ba^\dag\ba-\textbf{1}/2)(\bb + \bb^\dag)  - \chi(\ba^\dag\ba)(\bb^\dag\bb),
\label{eq:Hamiltonian}
\end{equation}
where $\alpha$ is the anharmonicity of the transmon qubit ($\ba$) and $\chi$ is its dispersive coupling to the readout cavity ($\bb$). The second term is the same as a resonant longitudinal interaction of strength $\zeta$ between the transmon qubit and the readout cavity. With this implementation, the interaction is gated: it can be instantly switched on/off, and is qubit-selective. We exploit this feature to multiplex several qubits with a single readout resonator and show that the target qubit can be measured non-destructively 98.4\% of the time, with minimal detrimental effects on the other qubits of the system.

\begin{figure}
\includegraphics{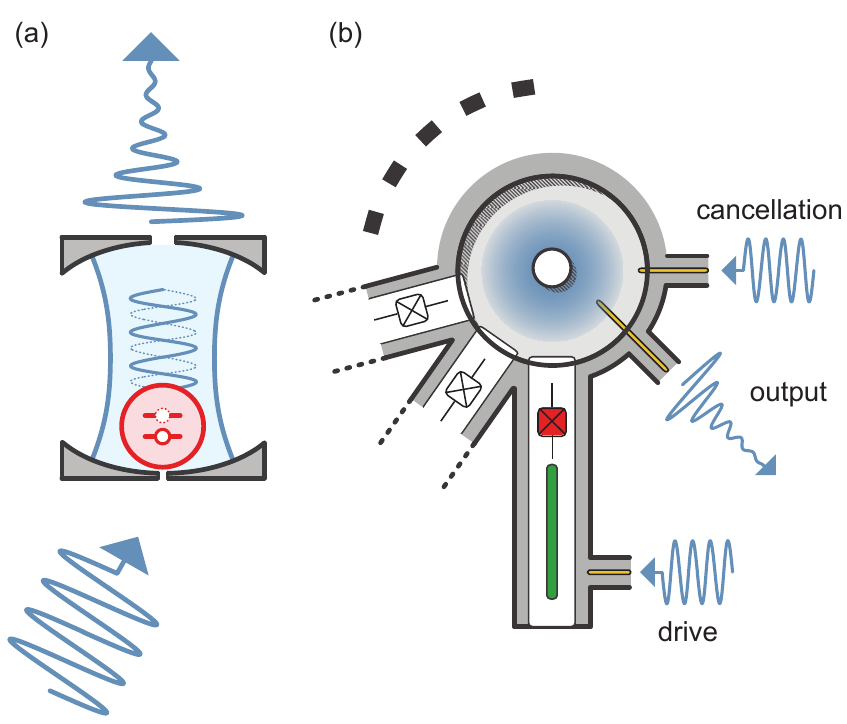}
\caption{\label{fig1}
(a) Principle of the experiment. A schematic superconducting two-level artificial atom (red) is placed where the field of a cavity (blue) is weak, for instance close to a partially transmitting mirror, to be in the weak coupling regime. When the atom is driven at the frequency of the cavity, the electromagnetic field of the cavity is spontaneously displaced with a sense which depends on the state of the atom (dotted or solid line) and exits through the main aperture. (b) Multi-qubit architecture. An arbitrary number of transmon chips are placed around the field of an aluminum post-cavity (three chips in the current experiment). The target transmon (red) is driven through a filter mode (green) which is coupled to a microwave input coupler. The number of photons in the cavity is kept minimal using a cancellation port (see text). The field of the cavity is measured using the strongly-coupled output port.
}
\end{figure}

The principle of the experiment is shown in Fig. \ref{fig1}(a). A superconducting qubit is weakly coupled to a low-Q microwave resonator such that their residual dispersive interaction $\chi$ is much smaller than the linewidth $\kappa$ \cite{Blais2004}. This so-called weak dispersive regime \cite{Schuster2007} is desirable since it mitigates the Purcell effect \cite{Houck2008}, the dephasing due to spurious thermal photons \cite{Gambetta2006, Sears2012}, and, more generally, any spurious coupling to other qubits through the cavity mode. However, the weak dispersive regime is usually unfavorable for qubit readout because it results in a slow measurement rate and, furthermore, requires populating the resonator with a large number of photons. This is a disadvantage in multi-qubit systems where photons in the shared resonator lead to unwanted decoherence in the qubits that are not being addressed. Here, we realize a fast readout while avoiding these drawbacks by implementing the aforementioned novel idea of conditional displacement readout. In our system a transmon is driven at the frequency of the cavity (Fig. \ref{fig1}(a)), resulting in the effective resonant longitudinal interaction. For a drive with an envelope of amplitude $\bar\epsilon(t)$ and a detuning $\Delta$ between the two modes, an analysis of the full Josephson Hamiltonian \cite{supp} gives

\begin{equation*}
\zeta(t) = \sqrt{2\alpha\chi}\frac{\bar\epsilon(t)}{\Delta}.
\end{equation*}
Since the strength $\zeta$ of this interaction depends on the product $\alpha\chi$, rather than $\chi$, it is possible to increase $\zeta$ while maintaining $\chi$ small. Thus, a fast readout is obtained while keeping the advantages of the weak dispersive coupling. 

We demonstrate these features using an aluminum cylindrical post-cavity \cite{Reagor2016} as a readout cavity ($\omega_c/2\pi =$ \SI{8.0}{\giga \hertz}) coupled to three transmon qubits (Fig. \ref{fig1}(b)). Our scheme is also compatible with a 2D architecture and a larger number of qubits. The target qubit ($\omega_q/2\pi =$ \SI{4.9}{\giga \hertz}) is coupled on one side to a stripline resonator ($\omega_f/2\pi =$ \SI{6.4}{\giga \hertz}), which is used as a filter mode with two roles. First, the filter mode is well-coupled both to the drive input pin (with coupling $\kappa_c/2\pi \approx $ \SI{8}{\kilo \hertz}) and to the qubit ($\chi_{qf}/2\pi = $ \SI{2.5}{\mega \hertz}) so that we can drive the qubit strongly off-resonance without limiting its coherence through the Purcell effect. Second, the presence of the filter mode increases the physical distance between the drive pin and the readout cavity and limits their direct coupling to much less than \SI{1}{\kilo \hertz}. To minimize the number of photons in the readout cavity introduced by this finite direct coupling, a phase-locked cancellation drive is applied to a cancellation port (with coupling $\kappa_c/2\pi \approx$ \SI{5}{\kilo \hertz}). Finally, the field is picked up by a strongly coupled output port which connects the cavity to a phase-sensitive amplifying chain \cite{Frattini2018} and to room temperature electronics \cite{supp}. We adjust the output coupling pin of the readout cavity in order to get an emission rate $\kappa =$ (\SI{100}{\nano \second})$^{-1}$, which sets the characteristic time of our measurement. The target qubit is characterized by an anharmonicity $\alpha/2\pi =$ \SI{221}{\mega \hertz} and is coupled to the cavity with a residual $\chi/2\pi \approx$ \SI{100}{\kilo \hertz} ($\chi \approx \kappa/16$). Two other qubits with similar parameters are coupled to the same cavity. We observe a range of qubit energy relaxation times $T_1$ between \SI{90}{\micro \second} - \SI{190}{\micro \second}, which vary, not atypically, from sample to sample. We present the data acquired for a qubit with $T_1 =$ \SI{90}{\micro \second}. The $T_2$-echo of our transmon varies between cooldowns in the range of \SI{30}{\micro \second} to \SI{170}{\micro \second}, for reasons which have not yet been pinned down, but which we believe to be independent from the effect we are demonstrating. 

\begin{figure*}
\includegraphics{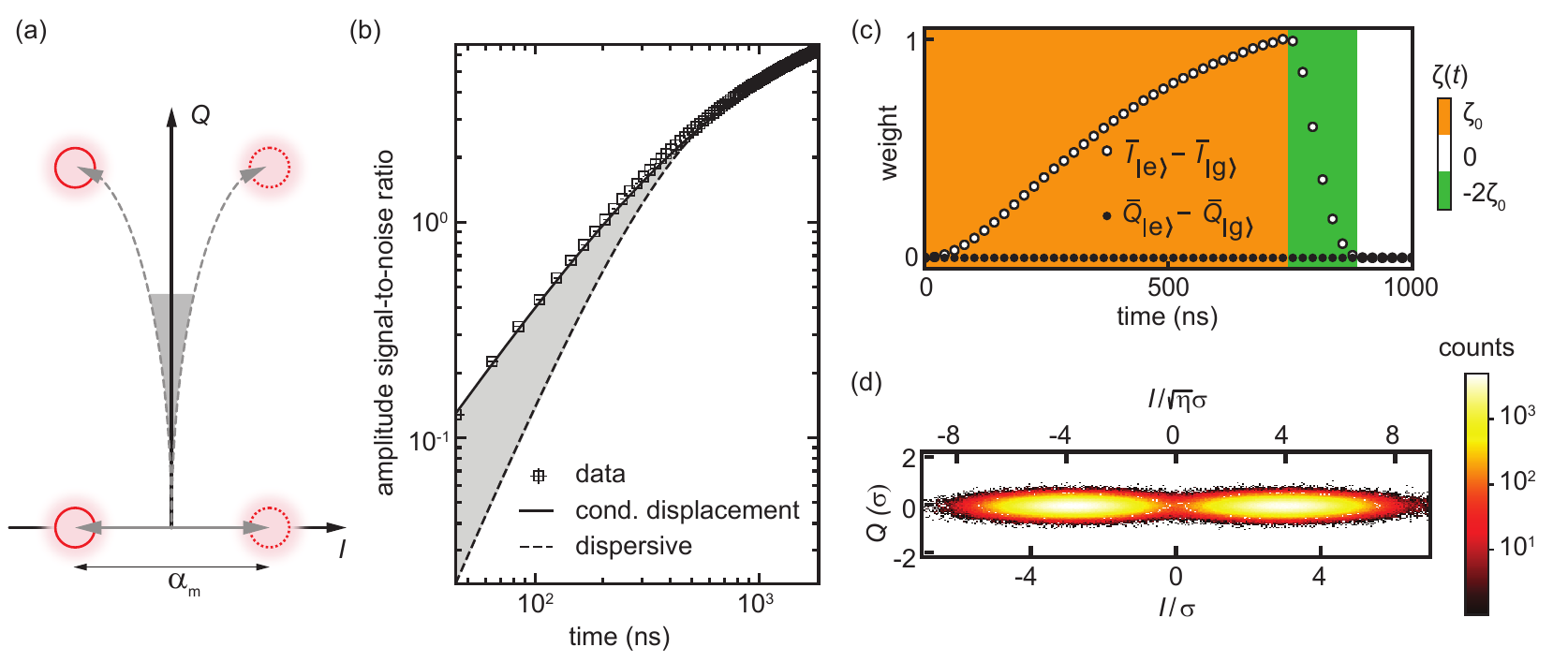}
\caption{\label{fig2}
Conditional coherent states separation. (a) Phase-space representation of this separation under an RF pulse implementing our engineered interaction (gray solid line) and under an RF pulse driving the cavity directly (gray dashed line).
In the former case, the field of the cavity is displaced along the \textit{I} quadrature with a sense that depends on the state of the transmon. The distance between the two possible steady states is noted $\alpha_m$. In the latter case, the cavity state would be displaced unconditionally along \textit{Q} and conditionally along \textit{I}. The gray area indicates that at small times, the two coherent states do not separate. (b) Log-Log plot of the amplitude SNR as a function of time. The data are fitted with the theoretical SNR for a conditional displacement (solid line), adjusted in amplitude with the efficiency $\eta$, determined independently, and with a fit parameter corresponding to the coupling strength. The dashed line corresponds to the optimal dispersive readout with the same parameters and $\chi=\kappa$. The gray area corresponds to the delay shown in (a). (c) Demodulation envelope comprising a depletion section. The coupling strength varies from $\zeta_0$ for \SI{750}{\nano \second} (orange), to $-2\zeta_0$ for \SI{120}{\nano \second} (green). (d) Histogram of the demodulated signal. The axis are calibrated using the calibrated efficiency $\eta$ and the width $\sigma$ of the Gaussians along the x-axis. The squeezing is due to the amplifier being phase-sensitive. }
\end{figure*}

In order to quantify the strength of the resonant longitudinal interaction, we turn on the drive for \SI{2}{\micro \second} and acquire $3\times 10^5$ trajectories with the target qubit initialized in $\ket{g}$ and $\ket{e}$. We use the ensemble average response for these two cases to determine the optimal demodulation envelope $(\bar{I}_{\ket{e}} - \bar{I}_{\ket{g}} - i(\bar{Q}_{\ket{e}} - \bar{Q}_{\ket{g}}))$ \cite{Gambetta2007, Ryan2015}, where the bar indicates the ensemble average. The optimal envelope is used to weigh single-shot trajectories and extract the SNR as a function of demodulation time. The SNR in amplitude, plotted in Fig. \ref{fig2}(b), is fit to the theoretical SNR for a conditional displacement demodulated with the optimal envelope \cite{supp}. The theory is only adjusted by an overall factor, which depends both on the efficiency $\eta = 0.6$ of the amplification chain, extracted independently \cite{supp,Bultink2018}, and the number $|\alpha_m|^2$ of measuring photons in steady-state (Fig. \ref{fig2}(a)). We estimate from the fit of the SNR $|\alpha_m|^2 = 2.6$ photons, which corresponds to a coupling strength $\zeta_0/2\pi = \alpha_m \kappa / 8\pi =$ \SI{1.28}{\mega \hertz} for the pulse of constant amplitude.

In the same plot, we compare the SNR of the displacement readout to the theoretical SNR of ideal dispersive readout with $\chi = \kappa$, using identical efficiency $\eta$ and photon number $|\alpha_m|^2$. In steady-state, by construction, the performance of both readouts converge to the same value $\propto(\kappa\tau)^{1/2}$. However, for the dispersive readout, the SNR grows much slower for initial times ($\kappa \tau \ll 1$). This can be understood from the initial cavity response, as shown in Fig. \ref{fig2}(a). For the dispersive readout, the cavity coherent state first rings up along the \textit{Q} quadrature at rate $\kappa$ and then separates, along the \textit{I} quadrature, at rate $\chi$, into the $\ket{g}$ and $\ket{e}$ components. On the other hand, for the conditional displacement readout, the two coherent states are displaced directly at rate $\kappa$. As the measurement is sensitive only to the separation, the conditional displacement readout is faster for short times.

The direct separation of the two coherent states along a single quadrature, as depicted in Fig. \ref{fig2}(a), is obtained for the optimal envelope, shown in Fig. \ref{fig2}(c) for a specific readout pulse length of \SI{750}{\nano \second}. By construction, the signal is contained within the \textit{I} quadrature and no response develops along \textit{Q}. Furthermore, to speed-up the measurement, we evacuate, near the end of the readout sequence, the cavity by reversing the amplitude of the pulse and hence, the strength of the coupling to $-2\zeta_0$ for \SI{120}{\nano \second}. A similar trick had been previously demonstrated for the dispersive case \cite{McClure2016, Bultink2018}. 

To quantify the discrimination power of the readout, we show in Fig. \ref{fig2}(d) the histogram corresponding to $1.5 \times 10^6$ single-shot measurements demodulated with the optimal envelope. 
The bottom x-axis is normalized by the apparent standard deviation of the two distributions, whereas the top x-axis is re-normalized with a factor $\sqrt{\eta}$ to depict the losses in the measurement chain. Since our setup uses a phase-sensitive amplifier to amplify along the \textit{I} quadrature, the distribution is squeezed along the \textit{Q} quadrature, which does not contain any information. The distributions along \textit{I} are separated by 5.8 standard deviations, corresponding to a discrimination power of 99.5\%.

\begin{figure}
\includegraphics{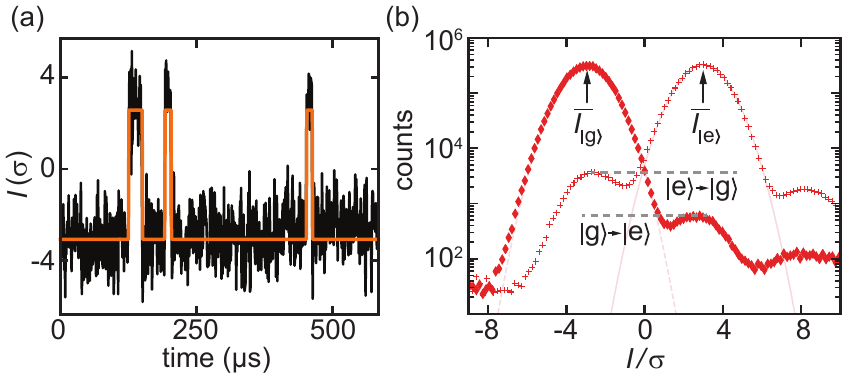}
\caption{\label{fig3} Quantum-non-demolition readout. (a) Results of successive single-shot measurements displaying 6 discrete jumps (black). The orange trace is a guide for the eye and is obtained with a latching filter applied to the data. The correlation between successive measurements indicates that the readout is non-destructive. (b) Histogram of the demodulated signal along the $I$ quadrature with post-selecting the qubit in $\ket{g}$ (diamonds) and in $\ket{e}$ (crosses). The two distributions are fitted with Gaussians (light red, dashed and solid lines). When the qubit starts in $\ket{g}$ (resp. $\ket{e}$) it mostly persists in $\ket{g}$ (resp. $\ket{e}$). The gray dashed lines emphasize the number of jumps from $\ket{g}$ (resp. $\ket{e}$) to $\ket{e}$ (resp. $\ket{g}$). 
}
\end{figure}

Although a good discrimination power is necessary, it is not sufficient to assess the overall merit of the readout. We further characterize the readout using two metrics: (1) the fidelity $\mathcal{F}$, which quantifies how accurately the measurement assigns the state prepared before the readout, and (2) the quantum-non-demolition metric $\mathcal{Q}$ (QND-ness), which quantifies how likely a qubit is to adopt its measured state after the readout. These metrics will be smaller than the discrimination power due to the qubit transitions during the readout, due themselves to either $T_1$ or induced by the drive. To estimate the two metrics, we perform a train of measurement pulses with no delay (Fig. \ref{fig3}(a)). The vast majority of measurement results are highly correlated with the previous one. Some rare measurement results display discrete transitions from one state to another. To estimate the fidelity, we plot in Fig. \ref{fig3}(b) the measurement distribution after a first stringent post-selection measurement: if the first measurement yields a value $I<\bar{I}_{\ket{g}}$ (resp. $I>\bar{I}_{\ket{e}}$), where the bar indicates the average of the distribution, we count the second measurement as being post-selected on $\ket{g}$ (resp. $\ket{e}$). We fit each distribution with a Gaussian and adjust a threshold to minimize the readout errors. We define the fidelities for the state $\ket{g}$ ($\ket{e}$) as $\mathcal{F}_g = 1 - p(g|e)$ ($\mathcal{F}_e = 1 - p(e|g)$), where $p(i|j)$ is the probability to measure the state $i$ if the qubit was initialized in $j$. We find $\mathcal{F}_g =$ 99.3\% and $\mathcal{F}_e =$ 98.5\%. From this, we define the total fidelity $\mathcal{F} = 1 - p(e|g) - p(g|e) =$ 97.8\%. On the other hand, the QND-ness is defined as $\mathcal{Q} = (p_{e, e} + p_{g, g})/2$, where $p_{i, i}$ is the probability to measure the state $i$ twice in two successive measurements. We find $\mathcal{Q} =$ 98.4\%. In practice $\mathcal{F}$ and $\mathcal{Q}$ are mainly limited by the energy relaxation of the qubit.

\begin{figure}
\includegraphics{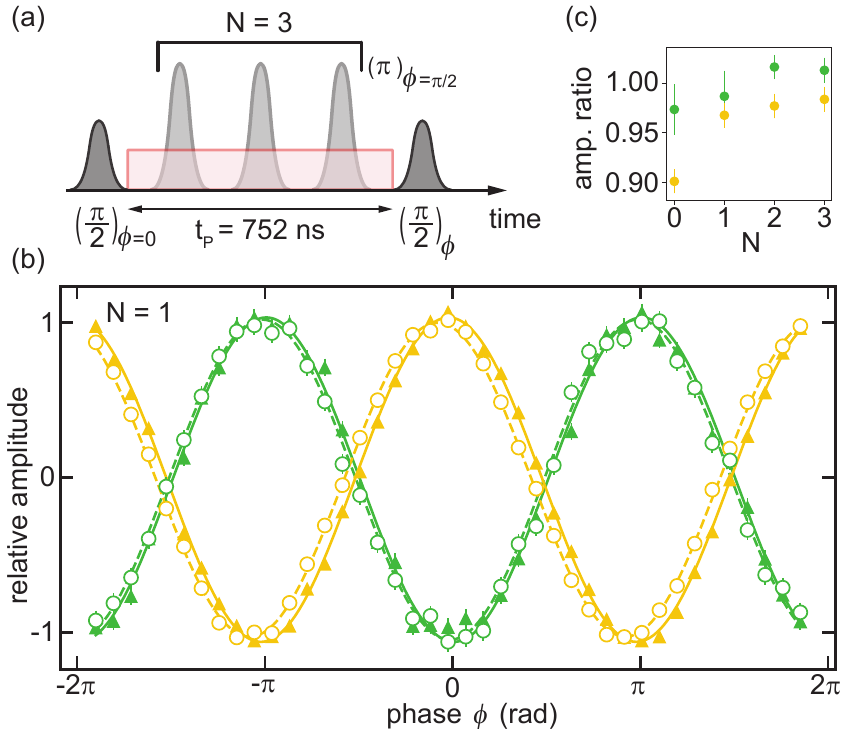}
\caption{\label{fig4} Coherence of two unmeasured qubits coupled to a common when the target qubit is measured. (a) Sequence for a Ramsey experiment with fixed length. Between two qubit rotations of $\pi/2$ (where $\phi$ indicates around what axis) either the measurement pulse (red) is applied or not. To mimic a decoupled quantum computation, N echo pulses are inserted between the two $\pi/2$ pulses. (b) Ramsey fringes while performing one echo pulse. The triangles and solid lines are respectively the data and fit for the control experiment. The circles and dashed lines are the data and fit when the target qubit is measured. The two curves are phase-shifted due to the Stark-shift. The y-axis is normalized to the contrast of the control experiment. The experiment is performed on two different qubits (yellow and green) and the data are shown with opposite phases for clarity. In (c) we show the evolution of the coherence for different numbers of echo pulses. The Ramsey contrasts are normalized by the amplitude of their respective control experiment. 
}
\end{figure}

Finally, we present how selective our measurement is in a multi-qubit architecture, which comprises 3 qubits coupled to the same readout resonator. The two qubits that are not targeted by the measurement have a dispersive coupling to the readout resonator that is similar to the dispersive coupling of the main qubit $\chi$ ($\ll \kappa$). As a consequence, when the target qubit is measured, the unmeasured qubits are dephased by the photon shot noise of the coherent state with a number of photons $\bar{n}_{tot} = |\alpha_m|^2/4$. The dephasing rate is $\Gamma_d = \bar{n}_{tot}(\chi/\kappa)^2\kappa$  and should be much smaller than the measurement rate of the target qubit. However, this dephasing is not inevitable since it can be mitigated by applying a dynamical decoupling sequence of pulses to the unmeasured qubits \cite{Viola1999, Lidar2014}. In fact, any realistic quantum computation on the unmeasured qubits would use such a dynamical decoupling sequence of pulses to mitigate environmental dephasing, which is often the main source of decoherence in cQED. Hence, the spurious dephasing due to the selective measurement will also be suppressed without having to adapt the pulse sequence on the unmeasured qubits. While measuring the target qubit, we assess the decoherence of the unmeasured qubits with a fixed-length Ramsey sequence with N interleaved $\pi$-pulses on the unmeasured qubits (Fig. \ref{fig4}(a)). As shown in Fig. \ref{fig4}(b), the Ramsey contrast for both unmeasured qubits is nearly independent of whether the target qubit is measured or not. In Fig. \ref{fig4}(c) we plot the ratio between the amplitudes in these two cases and observe that the measurement pulse adds at most 10\% of dephasing with no decoupling. Moreover, it is completely eliminated by inserting a few echo pulses in the Ramsey sequence. 

In conclusion, we have realized a new readout method for the state of superconducting qubits, in which the information of the qubit is coupled to a displacement along a single quadrature of a readout resonator. We have demonstrated fast and selective QND readout with this coupling in a multi-qubit architecture. This coupling is strong even when the dispersive shift is more than an order of magnitude smaller than the linewidth of the resonator, which can be beneficial to the coherence of the qubits. Our readout scheme can be made even faster with further optimizations of the system. More importantly, unlike the dispersive readout, our displacement readout can provide exponentially improved sensitivity by squeezing the microwave photons incident on the readout resonator \cite{Didier2015,Eddins2018}. The interaction we engineered is also useful beyond the readout of superconducting qubits. It can be applied to multi-qubit gates \cite{Kerman2013, Billangeon2015,Richer2016,Royer2017}, to the creation and correction of GKP codes \cite{Terhal2016} and pair-cat codes \cite{Albert2018}, and to single-photon \cite{Royer2018} and  photon-parity \cite{Puri2018} detection. 

We acknowledge U. Vool and P. Campagne-Ibarcq for helpful discussions. Facilities use was supported by the Yale SEAS clean room and the Yale Institute for Nanoscience and Quantum Engineering (YINQE). This research was supported by the Army Research Office (ARO) under Grants No. W911NF-14-1-0011, W911NF-18-1-0212 and W911NF-14-1-0563.

\bibliographystyle{apsrev4-1}
\bibliography{bib_Lreadout}

\begin{thebibliography}{40}%
\makeatletter
\providecommand \@ifxundefined [1]{%
 \@ifx{#1\undefined}
}%
\providecommand \@ifnum [1]{%
 \ifnum #1\expandafter \@firstoftwo
 \else \expandafter \@secondoftwo
 \fi
}%
\providecommand \@ifx [1]{%
 \ifx #1\expandafter \@firstoftwo
 \else \expandafter \@secondoftwo
 \fi
}%
\providecommand \natexlab [1]{#1}%
\providecommand \enquote  [1]{``#1''}%
\providecommand \bibnamefont  [1]{#1}%
\providecommand \bibfnamefont [1]{#1}%
\providecommand \citenamefont [1]{#1}%
\providecommand \href@noop [0]{\@secondoftwo}%
\providecommand \href [0]{\begingroup \@sanitize@url \@href}%
\providecommand \@href[1]{\@@startlink{#1}\@@href}%
\providecommand \@@href[1]{\endgroup#1\@@endlink}%
\providecommand \@sanitize@url [0]{\catcode `\\12\catcode `\$12\catcode
  `\&12\catcode `\#12\catcode `\^12\catcode `\_12\catcode `\%12\relax}%
\providecommand \@@startlink[1]{}%
\providecommand \@@endlink[0]{}%
\providecommand \url  [0]{\begingroup\@sanitize@url \@url }%
\providecommand \@url [1]{\endgroup\@href {#1}{\urlprefix }}%
\providecommand \urlprefix  [0]{URL }%
\providecommand \Eprint [0]{\href }%
\providecommand \doibase [0]{http://dx.doi.org/}%
\providecommand \selectlanguage [0]{\@gobble}%
\providecommand \bibinfo  [0]{\@secondoftwo}%
\providecommand \bibfield  [0]{\@secondoftwo}%
\providecommand \translation [1]{[#1]}%
\providecommand \BibitemOpen [0]{}%
\providecommand \bibitemStop [0]{}%
\providecommand \bibitemNoStop [0]{.\EOS\space}%
\providecommand \EOS [0]{\spacefactor3000\relax}%
\providecommand \BibitemShut  [1]{\csname bibitem#1\endcsname}%
\let\auto@bib@innerbib\@empty
\bibitem [{\citenamefont {DiVincenzo}(2000)}]{DiVincenzo2000}%
  \BibitemOpen
  \bibfield  {author} {\bibinfo {author} {\bibfnamefont {D.~P.}\ \bibnamefont
  {DiVincenzo}},\ }\href {\doibase
  10.1002/1521-3978(200009)48:9/11<771::AID-PROP771>3.0.CO;2-E} {\bibfield
  {journal} {\bibinfo  {journal} {Fortschritte der Physik}\ }\textbf {\bibinfo
  {volume} {48}},\ \bibinfo {pages} {771} (\bibinfo {year} {2000})}\BibitemShut
  {NoStop}%
\bibitem [{\citenamefont {Haroche}\ and\ \citenamefont
  {Raimond}(2006)}]{Haroche2006}%
  \BibitemOpen
  \bibfield  {author} {\bibinfo {author} {\bibfnamefont {S.}~\bibnamefont
  {Haroche}}\ and\ \bibinfo {author} {\bibfnamefont {J.}~\bibnamefont
  {Raimond}},\ }\href@noop {} {\emph {\bibinfo {title} {Exploring the Quantum:
  Atoms, Cavities and Photons.}}}\ (\bibinfo  {publisher} {Oxford University
  Press},\ \bibinfo {year} {2006})\BibitemShut {NoStop}%
\bibitem [{\citenamefont {Blais}\ \emph {et~al.}(2004)\citenamefont {Blais},
  \citenamefont {Huang}, \citenamefont {Wallraff}, \citenamefont {Girvin},\
  and\ \citenamefont {Schoelkopf}}]{Blais2004}%
  \BibitemOpen
  \bibfield  {author} {\bibinfo {author} {\bibfnamefont {A.}~\bibnamefont
  {Blais}}, \bibinfo {author} {\bibfnamefont {R.-S.}\ \bibnamefont {Huang}},
  \bibinfo {author} {\bibfnamefont {A.}~\bibnamefont {Wallraff}}, \bibinfo
  {author} {\bibfnamefont {S.~M.}\ \bibnamefont {Girvin}}, \ and\ \bibinfo
  {author} {\bibfnamefont {R.~J.}\ \bibnamefont {Schoelkopf}},\ }\href
  {\doibase 10.1103/PhysRevA.69.062320} {\bibfield  {journal} {\bibinfo
  {journal} {Phys. Rev. A}\ }\textbf {\bibinfo {volume} {69}},\ \bibinfo
  {pages} {62320} (\bibinfo {year} {2004})}\BibitemShut {NoStop}%
\bibitem [{\citenamefont {Wallraff}\ \emph {et~al.}(2004)\citenamefont
  {Wallraff}, \citenamefont {Schuster}, \citenamefont {Blais}, \citenamefont
  {Frunzio}, \citenamefont {Huang}, \citenamefont {Majer}, \citenamefont
  {Kumar}, \citenamefont {Girvin},\ and\ \citenamefont
  {Schoelkopf}}]{Wallraff2004}%
  \BibitemOpen
  \bibfield  {author} {\bibinfo {author} {\bibfnamefont {A.}~\bibnamefont
  {Wallraff}}, \bibinfo {author} {\bibfnamefont {D.~I.}\ \bibnamefont
  {Schuster}}, \bibinfo {author} {\bibfnamefont {A.}~\bibnamefont {Blais}},
  \bibinfo {author} {\bibfnamefont {L.}~\bibnamefont {Frunzio}}, \bibinfo
  {author} {\bibfnamefont {R.~S.}\ \bibnamefont {Huang}}, \bibinfo {author}
  {\bibfnamefont {J.}~\bibnamefont {Majer}}, \bibinfo {author} {\bibfnamefont
  {S.}~\bibnamefont {Kumar}}, \bibinfo {author} {\bibfnamefont {S.~M.}\
  \bibnamefont {Girvin}}, \ and\ \bibinfo {author} {\bibfnamefont {R.~J.}\
  \bibnamefont {Schoelkopf}},\ }\href@noop {} {\bibfield  {journal} {\bibinfo
  {journal} {Nature}\ }\textbf {\bibinfo {volume} {431}},\ \bibinfo {pages}
  {162} (\bibinfo {year} {2004})}\BibitemShut {NoStop}%
\bibitem [{\citenamefont {Gambetta}\ \emph {et~al.}(2008)\citenamefont
  {Gambetta}, \citenamefont {Blais}, \citenamefont {Boissonneault},
  \citenamefont {Houck}, \citenamefont {Schuster},\ and\ \citenamefont
  {Girvin}}]{Gambetta2008}%
  \BibitemOpen
  \bibfield  {author} {\bibinfo {author} {\bibfnamefont {J.}~\bibnamefont
  {Gambetta}}, \bibinfo {author} {\bibfnamefont {A.}~\bibnamefont {Blais}},
  \bibinfo {author} {\bibfnamefont {M.}~\bibnamefont {Boissonneault}}, \bibinfo
  {author} {\bibfnamefont {A.~A.}\ \bibnamefont {Houck}}, \bibinfo {author}
  {\bibfnamefont {D.~I.}\ \bibnamefont {Schuster}}, \ and\ \bibinfo {author}
  {\bibfnamefont {S.~M.}\ \bibnamefont {Girvin}},\ }\href@noop {} {\bibfield
  {journal} {\bibinfo  {journal} {Phys. Rev. A}\ }\textbf {\bibinfo {volume}
  {77}},\ \bibinfo {pages} {1} (\bibinfo {year} {2008})}\BibitemShut {NoStop}%
\bibitem [{\citenamefont {Picot}\ \emph {et~al.}(2008)\citenamefont {Picot},
  \citenamefont {Lupaşcu}, \citenamefont {Saito}, \citenamefont {Harmans},\
  and\ \citenamefont {Mooij}}]{Picot2008}%
  \BibitemOpen
  \bibfield  {author} {\bibinfo {author} {\bibfnamefont {T.}~\bibnamefont
  {Picot}}, \bibinfo {author} {\bibfnamefont {A.}~\bibnamefont {Lupaşcu}},
  \bibinfo {author} {\bibfnamefont {S.}~\bibnamefont {Saito}}, \bibinfo
  {author} {\bibfnamefont {C.~J. P.~M.}\ \bibnamefont {Harmans}}, \ and\
  \bibinfo {author} {\bibfnamefont {J.~E.}\ \bibnamefont {Mooij}},\ }\href
  {\doibase 10.1103/PhysRevB.78.132508} {\bibfield  {journal} {\bibinfo
  {journal} {Phys. Rev. B}\ }\textbf {\bibinfo {volume} {78}},\ \bibinfo
  {pages} {132508} (\bibinfo {year} {2008})}\BibitemShut {NoStop}%
\bibitem [{\citenamefont {Boissonneault}\ \emph {et~al.}(2009)\citenamefont
  {Boissonneault}, \citenamefont {Gambetta},\ and\ \citenamefont
  {Blais}}]{Boissonneault2009}%
  \BibitemOpen
  \bibfield  {author} {\bibinfo {author} {\bibfnamefont {M.}~\bibnamefont
  {Boissonneault}}, \bibinfo {author} {\bibfnamefont {J.~M.}\ \bibnamefont
  {Gambetta}}, \ and\ \bibinfo {author} {\bibfnamefont {A.}~\bibnamefont
  {Blais}},\ }\href@noop {} {\bibfield  {journal} {\bibinfo  {journal} {Phys.
  Rev. A}\ }\textbf {\bibinfo {volume} {79}},\ \bibinfo {pages} {13819}
  (\bibinfo {year} {2009})}\BibitemShut {NoStop}%
\bibitem [{\citenamefont {Slichter}\ \emph {et~al.}(2012)\citenamefont
  {Slichter}, \citenamefont {Vijay}, \citenamefont {Weber}, \citenamefont
  {Boutin}, \citenamefont {Boissonneault}, \citenamefont {Gambetta},
  \citenamefont {Blais},\ and\ \citenamefont {Siddiqi}}]{Slichter2012}%
  \BibitemOpen
  \bibfield  {author} {\bibinfo {author} {\bibfnamefont {D.~H.}\ \bibnamefont
  {Slichter}}, \bibinfo {author} {\bibfnamefont {R.}~\bibnamefont {Vijay}},
  \bibinfo {author} {\bibfnamefont {S.~J.}\ \bibnamefont {Weber}}, \bibinfo
  {author} {\bibfnamefont {S.}~\bibnamefont {Boutin}}, \bibinfo {author}
  {\bibfnamefont {M.}~\bibnamefont {Boissonneault}}, \bibinfo {author}
  {\bibfnamefont {J.~M.}\ \bibnamefont {Gambetta}}, \bibinfo {author}
  {\bibfnamefont {A.}~\bibnamefont {Blais}}, \ and\ \bibinfo {author}
  {\bibfnamefont {I.}~\bibnamefont {Siddiqi}},\ }\href {\doibase
  10.1103/PhysRevLett.109.153601} {\bibfield  {journal} {\bibinfo  {journal}
  {Phys. Rev. Lett.}\ }\textbf {\bibinfo {volume} {109}},\ \bibinfo {pages}
  {153601} (\bibinfo {year} {2012})}\BibitemShut {NoStop}%
\bibitem [{\citenamefont {Sank}\ \emph {et~al.}(2016)\citenamefont {Sank},
  \citenamefont {Chen}, \citenamefont {Khezri}, \citenamefont {Kelly},
  \citenamefont {Barends}, \citenamefont {Campbell}, \citenamefont {Chen},
  \citenamefont {Chiaro}, \citenamefont {Dunsworth}, \citenamefont {Fowler},
  \citenamefont {Jeffrey}, \citenamefont {Lucero}, \citenamefont {Megrant},
  \citenamefont {Mutus}, \citenamefont {Neeley}, \citenamefont {Neill},
  \citenamefont {O'Malley}, \citenamefont {Quintana}, \citenamefont {Roushan},
  \citenamefont {Vainsencher}, \citenamefont {White}, \citenamefont {Wenner},
  \citenamefont {Korotkov},\ and\ \citenamefont {Martinis}}]{Sank2016}%
  \BibitemOpen
  \bibfield  {author} {\bibinfo {author} {\bibfnamefont {D.}~\bibnamefont
  {Sank}}, \bibinfo {author} {\bibfnamefont {Z.}~\bibnamefont {Chen}}, \bibinfo
  {author} {\bibfnamefont {M.}~\bibnamefont {Khezri}}, \bibinfo {author}
  {\bibfnamefont {J.}~\bibnamefont {Kelly}}, \bibinfo {author} {\bibfnamefont
  {R.}~\bibnamefont {Barends}}, \bibinfo {author} {\bibfnamefont
  {B.}~\bibnamefont {Campbell}}, \bibinfo {author} {\bibfnamefont
  {Y.}~\bibnamefont {Chen}}, \bibinfo {author} {\bibfnamefont {B.}~\bibnamefont
  {Chiaro}}, \bibinfo {author} {\bibfnamefont {A.}~\bibnamefont {Dunsworth}},
  \bibinfo {author} {\bibfnamefont {A.}~\bibnamefont {Fowler}}, \bibinfo
  {author} {\bibfnamefont {E.}~\bibnamefont {Jeffrey}}, \bibinfo {author}
  {\bibfnamefont {E.}~\bibnamefont {Lucero}}, \bibinfo {author} {\bibfnamefont
  {A.}~\bibnamefont {Megrant}}, \bibinfo {author} {\bibfnamefont
  {J.}~\bibnamefont {Mutus}}, \bibinfo {author} {\bibfnamefont
  {M.}~\bibnamefont {Neeley}}, \bibinfo {author} {\bibfnamefont
  {C.}~\bibnamefont {Neill}}, \bibinfo {author} {\bibfnamefont {P.~J.~J.}\
  \bibnamefont {O'Malley}}, \bibinfo {author} {\bibfnamefont {C.}~\bibnamefont
  {Quintana}}, \bibinfo {author} {\bibfnamefont {P.}~\bibnamefont {Roushan}},
  \bibinfo {author} {\bibfnamefont {A.}~\bibnamefont {Vainsencher}}, \bibinfo
  {author} {\bibfnamefont {T.}~\bibnamefont {White}}, \bibinfo {author}
  {\bibfnamefont {J.}~\bibnamefont {Wenner}}, \bibinfo {author} {\bibfnamefont
  {A.~N.}\ \bibnamefont {Korotkov}}, \ and\ \bibinfo {author} {\bibfnamefont
  {J.~M.}\ \bibnamefont {Martinis}},\ }\href {\doibase
  10.1103/PhysRevLett.117.190503} {\bibfield  {journal} {\bibinfo  {journal}
  {Phys. Rev. Lett.}\ }\textbf {\bibinfo {volume} {117}},\ \bibinfo {pages}
  {190503} (\bibinfo {year} {2016})}\BibitemShut {NoStop}%
\bibitem [{\citenamefont {Kerman}(2013)}]{Kerman2013}%
  \BibitemOpen
  \bibfield  {author} {\bibinfo {author} {\bibfnamefont {A.~J.}\ \bibnamefont
  {Kerman}},\ }\href {\doibase 10.1088/1367-2630/15/12/123011} {\bibfield
  {journal} {\bibinfo  {journal} {New J. Phys.}\ }\textbf {\bibinfo {volume}
  {15}},\ \bibinfo {pages} {123011} (\bibinfo {year} {2013})}\BibitemShut
  {NoStop}%
\bibitem [{\citenamefont {Billangeon}\ \emph {et~al.}(2015)\citenamefont
  {Billangeon}, \citenamefont {Tsai},\ and\ \citenamefont
  {Nakamura}}]{Billangeon2015}%
  \BibitemOpen
  \bibfield  {author} {\bibinfo {author} {\bibfnamefont {P.~M.}\ \bibnamefont
  {Billangeon}}, \bibinfo {author} {\bibfnamefont {J.~S.}\ \bibnamefont
  {Tsai}}, \ and\ \bibinfo {author} {\bibfnamefont {Y.}~\bibnamefont
  {Nakamura}},\ }\href@noop {} {\bibfield  {journal} {\bibinfo  {journal}
  {Phys. Rev. B}\ }\textbf {\bibinfo {volume} {91}},\ \bibinfo {pages} {94517}
  (\bibinfo {year} {2015})}\BibitemShut {NoStop}%
\bibitem [{\citenamefont {Didier}\ \emph {et~al.}(2015)\citenamefont {Didier},
  \citenamefont {Bourassa},\ and\ \citenamefont {Blais}}]{Didier2015}%
  \BibitemOpen
  \bibfield  {author} {\bibinfo {author} {\bibfnamefont {N.}~\bibnamefont
  {Didier}}, \bibinfo {author} {\bibfnamefont {J.}~\bibnamefont {Bourassa}}, \
  and\ \bibinfo {author} {\bibfnamefont {A.}~\bibnamefont {Blais}},\
  }\href@noop {} {\bibfield  {journal} {\bibinfo  {journal} {Phys. Rev. Lett.}\
  }\textbf {\bibinfo {volume} {115}},\ \bibinfo {pages} {203601} (\bibinfo
  {year} {2015})}\BibitemShut {NoStop}%
\bibitem [{\citenamefont {Aspelmeyer}\ \emph {et~al.}(2014)\citenamefont
  {Aspelmeyer}, \citenamefont {Kippenberg},\ and\ \citenamefont
  {Marquardt}}]{Aspelmeyer2014}%
  \BibitemOpen
  \bibfield  {author} {\bibinfo {author} {\bibfnamefont {M.}~\bibnamefont
  {Aspelmeyer}}, \bibinfo {author} {\bibfnamefont {T.~J.}\ \bibnamefont
  {Kippenberg}}, \ and\ \bibinfo {author} {\bibfnamefont {F.}~\bibnamefont
  {Marquardt}},\ }\href@noop {} {\bibfield  {journal} {\bibinfo  {journal}
  {Rev. Mod. Phys.}\ }\textbf {\bibinfo {volume} {86}},\ \bibinfo {pages}
  {1391} (\bibinfo {year} {2014})}\BibitemShut {NoStop}%
\bibitem [{\citenamefont {Eddins}\ \emph {et~al.}(2018)\citenamefont {Eddins},
  \citenamefont {Schreppler}, \citenamefont {Toyli}, \citenamefont {Martin},
  \citenamefont {Hacohen-Gourgy}, \citenamefont {Govia}, \citenamefont
  {Ribeiro}, \citenamefont {Clerk},\ and\ \citenamefont
  {Siddiqi}}]{Eddins2018}%
  \BibitemOpen
  \bibfield  {author} {\bibinfo {author} {\bibfnamefont {A.}~\bibnamefont
  {Eddins}}, \bibinfo {author} {\bibfnamefont {S.}~\bibnamefont {Schreppler}},
  \bibinfo {author} {\bibfnamefont {D.~M.}\ \bibnamefont {Toyli}}, \bibinfo
  {author} {\bibfnamefont {L.~S.}\ \bibnamefont {Martin}}, \bibinfo {author}
  {\bibfnamefont {S.}~\bibnamefont {Hacohen-Gourgy}}, \bibinfo {author}
  {\bibfnamefont {L.~C.~G.}\ \bibnamefont {Govia}}, \bibinfo {author}
  {\bibfnamefont {H.}~\bibnamefont {Ribeiro}}, \bibinfo {author} {\bibfnamefont
  {A.~A.}\ \bibnamefont {Clerk}}, \ and\ \bibinfo {author} {\bibfnamefont
  {I.}~\bibnamefont {Siddiqi}},\ }\href@noop {} {\bibfield  {journal} {\bibinfo
   {journal} {Phys. Rev. Lett.}\ }\textbf {\bibinfo {volume} {120}},\ \bibinfo
  {pages} {040505} (\bibinfo {year} {2018})}\BibitemShut {NoStop}%
\bibitem [{\citenamefont {Viola}\ \emph {et~al.}(1999)\citenamefont {Viola},
  \citenamefont {Lloyd},\ and\ \citenamefont {Knill}}]{Viola1999}%
  \BibitemOpen
  \bibfield  {author} {\bibinfo {author} {\bibfnamefont {L.}~\bibnamefont
  {Viola}}, \bibinfo {author} {\bibfnamefont {S.}~\bibnamefont {Lloyd}}, \ and\
  \bibinfo {author} {\bibfnamefont {E.}~\bibnamefont {Knill}},\ }\href@noop {}
  {\bibfield  {journal} {\bibinfo  {journal} {Phys. Rev. Lett.}\ }\textbf
  {\bibinfo {volume} {83}},\ \bibinfo {pages} {4888} (\bibinfo {year}
  {1999})}\BibitemShut {NoStop}%
\bibitem [{\citenamefont {Clerk}\ \emph {et~al.}(2010)\citenamefont {Clerk},
  \citenamefont {Devoret}, \citenamefont {Girvin}, \citenamefont {Marquardt},\
  and\ \citenamefont {Schoelkopf}}]{Clerk2010}%
  \BibitemOpen
  \bibfield  {author} {\bibinfo {author} {\bibfnamefont {A.~A.}\ \bibnamefont
  {Clerk}}, \bibinfo {author} {\bibfnamefont {M.~H.}\ \bibnamefont {Devoret}},
  \bibinfo {author} {\bibfnamefont {S.~M.}\ \bibnamefont {Girvin}}, \bibinfo
  {author} {\bibfnamefont {F.}~\bibnamefont {Marquardt}}, \ and\ \bibinfo
  {author} {\bibfnamefont {R.~J.}\ \bibnamefont {Schoelkopf}},\ }\href@noop {}
  {\bibfield  {journal} {\bibinfo  {journal} {Rev. Mod. Phys.}\ }\textbf
  {\bibinfo {volume} {82}},\ \bibinfo {pages} {1155} (\bibinfo {year}
  {2010})}\BibitemShut {NoStop}%
\bibitem [{\citenamefont {Suh}\ \emph {et~al.}(2014)\citenamefont {Suh},
  \citenamefont {Weinstein}, \citenamefont {Lei}, \citenamefont {Wollman},
  \citenamefont {Steinke}, \citenamefont {Meystre}, \citenamefont {Clerk},\
  and\ \citenamefont {Schwab}}]{Suh2014}%
  \BibitemOpen
  \bibfield  {author} {\bibinfo {author} {\bibfnamefont {J.}~\bibnamefont
  {Suh}}, \bibinfo {author} {\bibfnamefont {A.~J.}\ \bibnamefont {Weinstein}},
  \bibinfo {author} {\bibfnamefont {C.~U.}\ \bibnamefont {Lei}}, \bibinfo
  {author} {\bibfnamefont {E.~E.}\ \bibnamefont {Wollman}}, \bibinfo {author}
  {\bibfnamefont {S.~K.}\ \bibnamefont {Steinke}}, \bibinfo {author}
  {\bibfnamefont {P.}~\bibnamefont {Meystre}}, \bibinfo {author} {\bibfnamefont
  {A.~A.}\ \bibnamefont {Clerk}}, \ and\ \bibinfo {author} {\bibfnamefont
  {K.~C.}\ \bibnamefont {Schwab}},\ }\href {\doibase 10.1126/science.1253258}
  {\bibfield  {journal} {\bibinfo  {journal} {Science}\ }\textbf {\bibinfo
  {volume} {344}},\ \bibinfo {pages} {1262} (\bibinfo {year}
  {2014})}\BibitemShut {NoStop}%
\bibitem [{\citenamefont {Vion}\ \emph {et~al.}(2002)\citenamefont {Vion},
  \citenamefont {Aassime}, \citenamefont {Cottet}, \citenamefont {Joyez},
  \citenamefont {Pothier}, \citenamefont {Urbina}, \citenamefont {Esteve},\
  and\ \citenamefont {Devoret}}]{Vion2002}%
  \BibitemOpen
  \bibfield  {author} {\bibinfo {author} {\bibfnamefont {D.}~\bibnamefont
  {Vion}}, \bibinfo {author} {\bibfnamefont {A.}~\bibnamefont {Aassime}},
  \bibinfo {author} {\bibfnamefont {A.}~\bibnamefont {Cottet}}, \bibinfo
  {author} {\bibfnamefont {P.}~\bibnamefont {Joyez}}, \bibinfo {author}
  {\bibfnamefont {H.}~\bibnamefont {Pothier}}, \bibinfo {author} {\bibfnamefont
  {C.}~\bibnamefont {Urbina}}, \bibinfo {author} {\bibfnamefont
  {D.}~\bibnamefont {Esteve}}, \ and\ \bibinfo {author} {\bibfnamefont {M.~H.}\
  \bibnamefont {Devoret}},\ }\href {\doibase 10.1126/science.1069372}
  {\bibfield  {journal} {\bibinfo  {journal} {Science}\ }\textbf {\bibinfo
  {volume} {296}},\ \bibinfo {pages} {886} (\bibinfo {year}
  {2002})}\BibitemShut {NoStop}%
\bibitem [{\citenamefont {Roy}\ \emph {et~al.}(2017)\citenamefont {Roy},
  \citenamefont {Kundu}, \citenamefont {Chand}, \citenamefont {Hazra},
  \citenamefont {Nehra}, \citenamefont {Cosmic}, \citenamefont {Ranadive},
  \citenamefont {Patankar}, \citenamefont {Damle},\ and\ \citenamefont
  {Vijay}}]{Roy2017}%
  \BibitemOpen
  \bibfield  {author} {\bibinfo {author} {\bibfnamefont {T.}~\bibnamefont
  {Roy}}, \bibinfo {author} {\bibfnamefont {S.}~\bibnamefont {Kundu}}, \bibinfo
  {author} {\bibfnamefont {M.}~\bibnamefont {Chand}}, \bibinfo {author}
  {\bibfnamefont {S.}~\bibnamefont {Hazra}}, \bibinfo {author} {\bibfnamefont
  {N.}~\bibnamefont {Nehra}}, \bibinfo {author} {\bibfnamefont
  {R.}~\bibnamefont {Cosmic}}, \bibinfo {author} {\bibfnamefont
  {A.}~\bibnamefont {Ranadive}}, \bibinfo {author} {\bibfnamefont {M.~P.}\
  \bibnamefont {Patankar}}, \bibinfo {author} {\bibfnamefont {K.}~\bibnamefont
  {Damle}}, \ and\ \bibinfo {author} {\bibfnamefont {R.}~\bibnamefont
  {Vijay}},\ }\href {\doibase 10.1103/PhysRevApplied.7.054025} {\bibfield
  {journal} {\bibinfo  {journal} {Phys. Rev. Applied}\ }\textbf {\bibinfo
  {volume} {7}},\ \bibinfo {pages} {054025} (\bibinfo {year}
  {2017})}\BibitemShut {NoStop}%
\bibitem [{\citenamefont {Eichler}\ and\ \citenamefont
  {Petta}(2018)}]{Eichler2018}%
  \BibitemOpen
  \bibfield  {author} {\bibinfo {author} {\bibfnamefont {C.}~\bibnamefont
  {Eichler}}\ and\ \bibinfo {author} {\bibfnamefont {J.~R.}\ \bibnamefont
  {Petta}},\ }\href {\doibase 10.1103/PhysRevLett.120.227702} {\bibfield
  {journal} {\bibinfo  {journal} {Phys. Rev. Lett.}\ }\textbf {\bibinfo
  {volume} {120}},\ \bibinfo {pages} {227702} (\bibinfo {year}
  {2018})}\BibitemShut {NoStop}%
\bibitem [{\citenamefont {Leghtas}\ \emph {et~al.}(2015)\citenamefont
  {Leghtas}, \citenamefont {Touzard}, \citenamefont {Pop}, \citenamefont {Kou},
  \citenamefont {Vlastakis}, \citenamefont {Petrenko}, \citenamefont {Sliwa},
  \citenamefont {Narla}, \citenamefont {Shankar}, \citenamefont {Hatridge},
  \citenamefont {Reagor}, \citenamefont {Frunzio}, \citenamefont {Schoelkopf},
  \citenamefont {Mirrahimi},\ and\ \citenamefont {Devoret}}]{Leghtas2015}%
  \BibitemOpen
  \bibfield  {author} {\bibinfo {author} {\bibfnamefont {Z.}~\bibnamefont
  {Leghtas}}, \bibinfo {author} {\bibfnamefont {S.}~\bibnamefont {Touzard}},
  \bibinfo {author} {\bibfnamefont {I.~M.}\ \bibnamefont {Pop}}, \bibinfo
  {author} {\bibfnamefont {A.}~\bibnamefont {Kou}}, \bibinfo {author}
  {\bibfnamefont {B.}~\bibnamefont {Vlastakis}}, \bibinfo {author}
  {\bibfnamefont {A.}~\bibnamefont {Petrenko}}, \bibinfo {author}
  {\bibfnamefont {K.~M.}\ \bibnamefont {Sliwa}}, \bibinfo {author}
  {\bibfnamefont {A.}~\bibnamefont {Narla}}, \bibinfo {author} {\bibfnamefont
  {S.}~\bibnamefont {Shankar}}, \bibinfo {author} {\bibfnamefont {M.~J.}\
  \bibnamefont {Hatridge}}, \bibinfo {author} {\bibfnamefont {M.}~\bibnamefont
  {Reagor}}, \bibinfo {author} {\bibfnamefont {L.}~\bibnamefont {Frunzio}},
  \bibinfo {author} {\bibfnamefont {R.~J.}\ \bibnamefont {Schoelkopf}},
  \bibinfo {author} {\bibfnamefont {M.}~\bibnamefont {Mirrahimi}}, \ and\
  \bibinfo {author} {\bibfnamefont {M.~H.}\ \bibnamefont {Devoret}},\ }\href
  {\doibase 10.1126/science.aaa2085} {\bibfield  {journal} {\bibinfo  {journal}
  {Science}\ }\textbf {\bibinfo {volume} {347}},\ \bibinfo {pages} {853}
  (\bibinfo {year} {2015})}\BibitemShut {NoStop}%
\bibitem [{sup()}]{supp}%
  \BibitemOpen
  \href@noop {} {}\bibinfo {note} {See supplementary materials for
  details}\BibitemShut {NoStop}%
\bibitem [{\citenamefont {Schuster}\ \emph {et~al.}(2007)\citenamefont
  {Schuster}, \citenamefont {Houck}, \citenamefont {Schreier}, \citenamefont
  {Wallraff}, \citenamefont {Gambetta}, \citenamefont {Blais}, \citenamefont
  {Frunzio}, \citenamefont {Majer}, \citenamefont {Johnson}, \citenamefont
  {Devoret}, \citenamefont {Girvin},\ and\ \citenamefont
  {Schoelkopf}}]{Schuster2007}%
  \BibitemOpen
  \bibfield  {author} {\bibinfo {author} {\bibfnamefont {D.~I.}\ \bibnamefont
  {Schuster}}, \bibinfo {author} {\bibfnamefont {A.~A.}\ \bibnamefont {Houck}},
  \bibinfo {author} {\bibfnamefont {J.~A.}\ \bibnamefont {Schreier}}, \bibinfo
  {author} {\bibfnamefont {A.}~\bibnamefont {Wallraff}}, \bibinfo {author}
  {\bibfnamefont {J.~M.}\ \bibnamefont {Gambetta}}, \bibinfo {author}
  {\bibfnamefont {A.}~\bibnamefont {Blais}}, \bibinfo {author} {\bibfnamefont
  {L.}~\bibnamefont {Frunzio}}, \bibinfo {author} {\bibfnamefont
  {J.}~\bibnamefont {Majer}}, \bibinfo {author} {\bibfnamefont
  {B.}~\bibnamefont {Johnson}}, \bibinfo {author} {\bibfnamefont {M.~H.}\
  \bibnamefont {Devoret}}, \bibinfo {author} {\bibfnamefont {S.~M.}\
  \bibnamefont {Girvin}}, \ and\ \bibinfo {author} {\bibfnamefont {R.~J.}\
  \bibnamefont {Schoelkopf}},\ }\href {\doibase 10.1038/nature05461} {\bibfield
   {journal} {\bibinfo  {journal} {Nature}\ }\textbf {\bibinfo {volume}
  {445}},\ \bibinfo {pages} {515} (\bibinfo {year} {2007})}\BibitemShut
  {NoStop}%
\bibitem [{\citenamefont {Houck}\ \emph {et~al.}(2008)\citenamefont {Houck},
  \citenamefont {Schreier}, \citenamefont {Johnson}, \citenamefont {Chow},
  \citenamefont {Koch}, \citenamefont {Gambetta}, \citenamefont {Schuster},
  \citenamefont {Frunzio}, \citenamefont {Devoret}, \citenamefont {Girvin},\
  and\ \citenamefont {Schoelkopf}}]{Houck2008}%
  \BibitemOpen
  \bibfield  {author} {\bibinfo {author} {\bibfnamefont {A.~A.}\ \bibnamefont
  {Houck}}, \bibinfo {author} {\bibfnamefont {J.~A.}\ \bibnamefont {Schreier}},
  \bibinfo {author} {\bibfnamefont {B.~R.}\ \bibnamefont {Johnson}}, \bibinfo
  {author} {\bibfnamefont {J.~M.}\ \bibnamefont {Chow}}, \bibinfo {author}
  {\bibfnamefont {J.}~\bibnamefont {Koch}}, \bibinfo {author} {\bibfnamefont
  {J.~M.}\ \bibnamefont {Gambetta}}, \bibinfo {author} {\bibfnamefont {D.~I.}\
  \bibnamefont {Schuster}}, \bibinfo {author} {\bibfnamefont {L.}~\bibnamefont
  {Frunzio}}, \bibinfo {author} {\bibfnamefont {M.~H.}\ \bibnamefont
  {Devoret}}, \bibinfo {author} {\bibfnamefont {S.~M.}\ \bibnamefont {Girvin}},
  \ and\ \bibinfo {author} {\bibfnamefont {R.~J.}\ \bibnamefont {Schoelkopf}},\
  }\href {\doibase 10.1103/PhysRevLett.101.080502} {\bibfield  {journal}
  {\bibinfo  {journal} {Phys. Rev. Lett.}\ }\textbf {\bibinfo {volume} {101}},\
  \bibinfo {pages} {080502} (\bibinfo {year} {2008})}\BibitemShut {NoStop}%
\bibitem [{\citenamefont {Gambetta}\ \emph {et~al.}(2006)\citenamefont
  {Gambetta}, \citenamefont {Blais}, \citenamefont {Schuster}, \citenamefont
  {Wallraff}, \citenamefont {Frunzio}, \citenamefont {Majer}, \citenamefont
  {Devoret}, \citenamefont {Girvin},\ and\ \citenamefont
  {Schoelkopf}}]{Gambetta2006}%
  \BibitemOpen
  \bibfield  {author} {\bibinfo {author} {\bibfnamefont {J.}~\bibnamefont
  {Gambetta}}, \bibinfo {author} {\bibfnamefont {A.}~\bibnamefont {Blais}},
  \bibinfo {author} {\bibfnamefont {D.~I.}\ \bibnamefont {Schuster}}, \bibinfo
  {author} {\bibfnamefont {A.}~\bibnamefont {Wallraff}}, \bibinfo {author}
  {\bibfnamefont {L.}~\bibnamefont {Frunzio}}, \bibinfo {author} {\bibfnamefont
  {J.}~\bibnamefont {Majer}}, \bibinfo {author} {\bibfnamefont {M.~H.}\
  \bibnamefont {Devoret}}, \bibinfo {author} {\bibfnamefont {S.~M.}\
  \bibnamefont {Girvin}}, \ and\ \bibinfo {author} {\bibfnamefont {R.~J.}\
  \bibnamefont {Schoelkopf}},\ }\href@noop {} {\bibfield  {journal} {\bibinfo
  {journal} {Phys. Rev. A}\ }\textbf {\bibinfo {volume} {74}},\ \bibinfo
  {pages} {042318} (\bibinfo {year} {2006})}\BibitemShut {NoStop}%
\bibitem [{\citenamefont {Sears}\ \emph {et~al.}(2012)\citenamefont {Sears},
  \citenamefont {Petrenko}, \citenamefont {Catelani}, \citenamefont {Sun},
  \citenamefont {Paik}, \citenamefont {Kirchmair}, \citenamefont {Frunzio},
  \citenamefont {Glazman}, \citenamefont {Girvin},\ and\ \citenamefont
  {Schoelkopf}}]{Sears2012}%
  \BibitemOpen
  \bibfield  {author} {\bibinfo {author} {\bibfnamefont {A.~P.}\ \bibnamefont
  {Sears}}, \bibinfo {author} {\bibfnamefont {A.}~\bibnamefont {Petrenko}},
  \bibinfo {author} {\bibfnamefont {G.}~\bibnamefont {Catelani}}, \bibinfo
  {author} {\bibfnamefont {L.}~\bibnamefont {Sun}}, \bibinfo {author}
  {\bibfnamefont {H.}~\bibnamefont {Paik}}, \bibinfo {author} {\bibfnamefont
  {G.}~\bibnamefont {Kirchmair}}, \bibinfo {author} {\bibfnamefont
  {L.}~\bibnamefont {Frunzio}}, \bibinfo {author} {\bibfnamefont {L.~I.}\
  \bibnamefont {Glazman}}, \bibinfo {author} {\bibfnamefont {S.~M.}\
  \bibnamefont {Girvin}}, \ and\ \bibinfo {author} {\bibfnamefont {R.~J.}\
  \bibnamefont {Schoelkopf}},\ }\href@noop {} {\bibfield  {journal} {\bibinfo
  {journal} {Phys. Rev. B}\ }\textbf {\bibinfo {volume} {86}},\ \bibinfo
  {pages} {1} (\bibinfo {year} {2012})}\BibitemShut {NoStop}%
\bibitem [{\citenamefont {Reagor}\ \emph {et~al.}(2016)\citenamefont {Reagor},
  \citenamefont {Pfaff}, \citenamefont {Axline}, \citenamefont {Heeres},
  \citenamefont {Ofek}, \citenamefont {Sliwa}, \citenamefont {Holland},
  \citenamefont {Wang}, \citenamefont {Blumoff}, \citenamefont {Chou},
  \citenamefont {Hatridge}, \citenamefont {Frunzio}, \citenamefont {Devoret},
  \citenamefont {Jiang},\ and\ \citenamefont {Schoelkopf}}]{Reagor2016}%
  \BibitemOpen
  \bibfield  {author} {\bibinfo {author} {\bibfnamefont {M.}~\bibnamefont
  {Reagor}}, \bibinfo {author} {\bibfnamefont {W.}~\bibnamefont {Pfaff}},
  \bibinfo {author} {\bibfnamefont {C.}~\bibnamefont {Axline}}, \bibinfo
  {author} {\bibfnamefont {R.~W.}\ \bibnamefont {Heeres}}, \bibinfo {author}
  {\bibfnamefont {N.}~\bibnamefont {Ofek}}, \bibinfo {author} {\bibfnamefont
  {K.}~\bibnamefont {Sliwa}}, \bibinfo {author} {\bibfnamefont
  {E.}~\bibnamefont {Holland}}, \bibinfo {author} {\bibfnamefont
  {C.}~\bibnamefont {Wang}}, \bibinfo {author} {\bibfnamefont {J.}~\bibnamefont
  {Blumoff}}, \bibinfo {author} {\bibfnamefont {K.}~\bibnamefont {Chou}},
  \bibinfo {author} {\bibfnamefont {M.~J.}\ \bibnamefont {Hatridge}}, \bibinfo
  {author} {\bibfnamefont {L.}~\bibnamefont {Frunzio}}, \bibinfo {author}
  {\bibfnamefont {M.~H.}\ \bibnamefont {Devoret}}, \bibinfo {author}
  {\bibfnamefont {L.}~\bibnamefont {Jiang}}, \ and\ \bibinfo {author}
  {\bibfnamefont {R.~J.}\ \bibnamefont {Schoelkopf}},\ }\href@noop {}
  {\bibfield  {journal} {\bibinfo  {journal} {Phys. Rev. B}\ }\textbf {\bibinfo
  {volume} {94}},\ \bibinfo {pages} {14506} (\bibinfo {year}
  {2016})}\BibitemShut {NoStop}%
\bibitem [{\citenamefont {Frattini}\ \emph {et~al.}(2018)\citenamefont
  {Frattini}, \citenamefont {Sivak}, \citenamefont {Lingenfelter},
  \citenamefont {Shankar},\ and\ \citenamefont {Devoret}}]{Frattini2018}%
  \BibitemOpen
  \bibfield  {author} {\bibinfo {author} {\bibfnamefont {N.~E.}\ \bibnamefont
  {Frattini}}, \bibinfo {author} {\bibfnamefont {V.~V.}\ \bibnamefont {Sivak}},
  \bibinfo {author} {\bibfnamefont {A.}~\bibnamefont {Lingenfelter}}, \bibinfo
  {author} {\bibfnamefont {S.}~\bibnamefont {Shankar}}, \ and\ \bibinfo
  {author} {\bibfnamefont {M.~H.}\ \bibnamefont {Devoret}},\ }\href
  {http://arxiv.org/abs/1806.06093} {\bibfield  {journal} {\bibinfo  {journal}
  {arXiv:1806.06093}\ } (\bibinfo {year} {2018})}\BibitemShut {NoStop}%
\bibitem [{\citenamefont {Gambetta}\ \emph {et~al.}(2007)\citenamefont
  {Gambetta}, \citenamefont {Braff}, \citenamefont {Wallraff}, \citenamefont
  {Girvin},\ and\ \citenamefont {Schoelkopf}}]{Gambetta2007}%
  \BibitemOpen
  \bibfield  {author} {\bibinfo {author} {\bibfnamefont {J.}~\bibnamefont
  {Gambetta}}, \bibinfo {author} {\bibfnamefont {W.~A.}\ \bibnamefont {Braff}},
  \bibinfo {author} {\bibfnamefont {A.}~\bibnamefont {Wallraff}}, \bibinfo
  {author} {\bibfnamefont {S.~M.}\ \bibnamefont {Girvin}}, \ and\ \bibinfo
  {author} {\bibfnamefont {R.~J.}\ \bibnamefont {Schoelkopf}},\ }\href
  {\doibase 10.1103/PhysRevA.76.012325} {\bibfield  {journal} {\bibinfo
  {journal} {Phys. Rev. A}\ }\textbf {\bibinfo {volume} {76}},\ \bibinfo
  {pages} {012325} (\bibinfo {year} {2007})}\BibitemShut {NoStop}%
\bibitem [{\citenamefont {Ryan}\ \emph {et~al.}(2015)\citenamefont {Ryan},
  \citenamefont {Johnson}, \citenamefont {Gambetta}, \citenamefont {Chow},
  \citenamefont {da~Silva}, \citenamefont {Dial},\ and\ \citenamefont
  {Ohki}}]{Ryan2015}%
  \BibitemOpen
  \bibfield  {author} {\bibinfo {author} {\bibfnamefont {C.~A.}\ \bibnamefont
  {Ryan}}, \bibinfo {author} {\bibfnamefont {B.~R.}\ \bibnamefont {Johnson}},
  \bibinfo {author} {\bibfnamefont {J.~M.}\ \bibnamefont {Gambetta}}, \bibinfo
  {author} {\bibfnamefont {J.~M.}\ \bibnamefont {Chow}}, \bibinfo {author}
  {\bibfnamefont {M.~P.}\ \bibnamefont {da~Silva}}, \bibinfo {author}
  {\bibfnamefont {O.~E.}\ \bibnamefont {Dial}}, \ and\ \bibinfo {author}
  {\bibfnamefont {T.~A.}\ \bibnamefont {Ohki}},\ }\href {\doibase
  10.1103/PhysRevA.91.022118} {\bibfield  {journal} {\bibinfo  {journal} {Phys.
  Rev. A}\ }\textbf {\bibinfo {volume} {91}},\ \bibinfo {pages} {022118}
  (\bibinfo {year} {2015})}\BibitemShut {NoStop}%
\bibitem [{\citenamefont {Bultink}\ \emph {et~al.}(2018)\citenamefont
  {Bultink}, \citenamefont {Tarasinski}, \citenamefont {Haandb{\ae}k},
  \citenamefont {Poletto}, \citenamefont {Haider}, \citenamefont {Michalak},
  \citenamefont {Bruno},\ and\ \citenamefont {DiCarlo}}]{Bultink2018}%
  \BibitemOpen
  \bibfield  {author} {\bibinfo {author} {\bibfnamefont {C.~C.}\ \bibnamefont
  {Bultink}}, \bibinfo {author} {\bibfnamefont {B.}~\bibnamefont {Tarasinski}},
  \bibinfo {author} {\bibfnamefont {N.}~\bibnamefont {Haandb{\ae}k}}, \bibinfo
  {author} {\bibfnamefont {S.}~\bibnamefont {Poletto}}, \bibinfo {author}
  {\bibfnamefont {N.}~\bibnamefont {Haider}}, \bibinfo {author} {\bibfnamefont
  {D.~J.}\ \bibnamefont {Michalak}}, \bibinfo {author} {\bibfnamefont
  {A.}~\bibnamefont {Bruno}}, \ and\ \bibinfo {author} {\bibfnamefont
  {L.}~\bibnamefont {DiCarlo}},\ }\href {\doibase 10.1063/1.5015954} {\bibfield
   {journal} {\bibinfo  {journal} {Appl. Phys. Lett.}\ }\textbf {\bibinfo
  {volume} {112}},\ \bibinfo {pages} {092601} (\bibinfo {year}
  {2018})}\BibitemShut {NoStop}%
\bibitem [{\citenamefont {McClure}\ \emph {et~al.}(2016)\citenamefont
  {McClure}, \citenamefont {Paik}, \citenamefont {Bishop}, \citenamefont
  {Steffen}, \citenamefont {Chow},\ and\ \citenamefont
  {Gambetta}}]{McClure2016}%
  \BibitemOpen
  \bibfield  {author} {\bibinfo {author} {\bibfnamefont {D.~T.}\ \bibnamefont
  {McClure}}, \bibinfo {author} {\bibfnamefont {H.}~\bibnamefont {Paik}},
  \bibinfo {author} {\bibfnamefont {L.~S.}\ \bibnamefont {Bishop}}, \bibinfo
  {author} {\bibfnamefont {M.}~\bibnamefont {Steffen}}, \bibinfo {author}
  {\bibfnamefont {J.~M.}\ \bibnamefont {Chow}}, \ and\ \bibinfo {author}
  {\bibfnamefont {J.~M.}\ \bibnamefont {Gambetta}},\ }\href {\doibase
  10.1103/PhysRevApplied.5.011001} {\bibfield  {journal} {\bibinfo  {journal}
  {Phys. Rev. Applied}\ }\textbf {\bibinfo {volume} {5}},\ \bibinfo {pages}
  {011001} (\bibinfo {year} {2016})}\BibitemShut {NoStop}%
\bibitem [{\citenamefont {Lidar}(2014)}]{Lidar2014}%
  \BibitemOpen
  \bibfield  {author} {\bibinfo {author} {\bibfnamefont {D.~A.}\ \bibnamefont
  {Lidar}},\ }\href {\doibase 10.1002/9781118742631.ch11} {\bibfield  {journal}
  {\bibinfo  {journal} {Adv. Chem. Phys.}\ }\textbf {\bibinfo {volume} {154}},\
  \bibinfo {pages} {295} (\bibinfo {year} {2014})}\BibitemShut {NoStop}%
\bibitem [{\citenamefont {Richer}\ and\ \citenamefont
  {Divincenzo}(2016)}]{Richer2016}%
  \BibitemOpen
  \bibfield  {author} {\bibinfo {author} {\bibfnamefont {S.}~\bibnamefont
  {Richer}}\ and\ \bibinfo {author} {\bibfnamefont {D.}~\bibnamefont
  {Divincenzo}},\ }\href@noop {} {\bibfield  {journal} {\bibinfo  {journal}
  {Phys. Rev. B}\ }\textbf {\bibinfo {volume} {93}} (\bibinfo {year}
  {2016})}\BibitemShut {NoStop}%
\bibitem [{\citenamefont {Royer}\ \emph {et~al.}(2017)\citenamefont {Royer},
  \citenamefont {Grimsmo}, \citenamefont {Didier},\ and\ \citenamefont
  {Blais}}]{Royer2017}%
  \BibitemOpen
  \bibfield  {author} {\bibinfo {author} {\bibfnamefont {B.}~\bibnamefont
  {Royer}}, \bibinfo {author} {\bibfnamefont {A.~L.}\ \bibnamefont {Grimsmo}},
  \bibinfo {author} {\bibfnamefont {N.}~\bibnamefont {Didier}}, \ and\ \bibinfo
  {author} {\bibfnamefont {A.}~\bibnamefont {Blais}},\ }\href@noop {}
  {\bibfield  {journal} {\bibinfo  {journal} {Quantum}\ }\textbf {\bibinfo
  {volume} {1}},\ \bibinfo {pages} {11} (\bibinfo {year} {2017})}\BibitemShut
  {NoStop}%
\bibitem [{\citenamefont {Terhal}\ and\ \citenamefont
  {Weigand}(2016)}]{Terhal2016}%
  \BibitemOpen
  \bibfield  {author} {\bibinfo {author} {\bibfnamefont {B.~M.}\ \bibnamefont
  {Terhal}}\ and\ \bibinfo {author} {\bibfnamefont {D.}~\bibnamefont
  {Weigand}},\ }\href@noop {} {\bibfield  {journal} {\bibinfo  {journal} {Phys.
  Rev. A}\ }\textbf {\bibinfo {volume} {93}},\ \bibinfo {pages} {012315}
  (\bibinfo {year} {2016})}\BibitemShut {NoStop}%
\bibitem [{\citenamefont {Albert}\ \emph {et~al.}(2018)\citenamefont {Albert},
  \citenamefont {Mundhada}, \citenamefont {Grimm}, \citenamefont {Touzard},
  \citenamefont {Devoret},\ and\ \citenamefont {Jiang}}]{Albert2018}%
  \BibitemOpen
  \bibfield  {author} {\bibinfo {author} {\bibfnamefont {V.~V.}\ \bibnamefont
  {Albert}}, \bibinfo {author} {\bibfnamefont {S.~O.}\ \bibnamefont
  {Mundhada}}, \bibinfo {author} {\bibfnamefont {A.}~\bibnamefont {Grimm}},
  \bibinfo {author} {\bibfnamefont {S.}~\bibnamefont {Touzard}}, \bibinfo
  {author} {\bibfnamefont {M.~H.}\ \bibnamefont {Devoret}}, \ and\ \bibinfo
  {author} {\bibfnamefont {L.}~\bibnamefont {Jiang}},\ }\href@noop {}
  {\bibfield  {journal} {\bibinfo  {journal} {arXiv:1801.05897}\ } (\bibinfo
  {year} {2018})}\BibitemShut {NoStop}%
\bibitem [{\citenamefont {Royer}\ \emph {et~al.}(2018)\citenamefont {Royer},
  \citenamefont {Grimsmo}, \citenamefont {Choquette-Poitevin},\ and\
  \citenamefont {Blais}}]{Royer2018}%
  \BibitemOpen
  \bibfield  {author} {\bibinfo {author} {\bibfnamefont {B.}~\bibnamefont
  {Royer}}, \bibinfo {author} {\bibfnamefont {A.~L.}\ \bibnamefont {Grimsmo}},
  \bibinfo {author} {\bibfnamefont {A.}~\bibnamefont {Choquette-Poitevin}}, \
  and\ \bibinfo {author} {\bibfnamefont {A.}~\bibnamefont {Blais}},\ }\href
  {\doibase 10.1103/PhysRevLett.120.203602} {\bibfield  {journal} {\bibinfo
  {journal} {Phys. Rev. Lett.}\ }\textbf {\bibinfo {volume} {120}},\ \bibinfo
  {pages} {203602} (\bibinfo {year} {2018})}\BibitemShut {NoStop}%
\bibitem [{\citenamefont {Puri}\ \emph {et~al.}(2018)\citenamefont {Puri},
  \citenamefont {Grimm}, \citenamefont {Campagne-Ibarcq}, \citenamefont
  {Eickbusch}, \citenamefont {Noh}, \citenamefont {Roberts}, \citenamefont
  {Jiang}, \citenamefont {Mirrahimi}, \citenamefont {Devoret},\ and\
  \citenamefont {Girvin}}]{Puri2018}%
  \BibitemOpen
  \bibfield  {author} {\bibinfo {author} {\bibfnamefont {S.}~\bibnamefont
  {Puri}}, \bibinfo {author} {\bibfnamefont {A.}~\bibnamefont {Grimm}},
  \bibinfo {author} {\bibfnamefont {P.}~\bibnamefont {Campagne-Ibarcq}},
  \bibinfo {author} {\bibfnamefont {A.}~\bibnamefont {Eickbusch}}, \bibinfo
  {author} {\bibfnamefont {K.}~\bibnamefont {Noh}}, \bibinfo {author}
  {\bibfnamefont {G.}~\bibnamefont {Roberts}}, \bibinfo {author} {\bibfnamefont
  {L.}~\bibnamefont {Jiang}}, \bibinfo {author} {\bibfnamefont
  {M.}~\bibnamefont {Mirrahimi}}, \bibinfo {author} {\bibfnamefont {M.~H.}\
  \bibnamefont {Devoret}}, \ and\ \bibinfo {author} {\bibfnamefont {S.~M.}\
  \bibnamefont {Girvin}},\ }\href@noop {} {\bibfield  {journal} {\bibinfo
  {journal} {arXiv:1807.09334}\ } (\bibinfo {year} {2018})}\BibitemShut
  {NoStop}%
\bibitem [{\citenamefont {Blumoff}\ \emph {et~al.}(2016)\citenamefont
  {Blumoff}, \citenamefont {Chou}, \citenamefont {Shen}, \citenamefont
  {Reagor}, \citenamefont {Axline}, \citenamefont {Brierley}, \citenamefont
  {Silveri}, \citenamefont {Wang}, \citenamefont {Vlastakis}, \citenamefont
  {Nigg}, \citenamefont {Frunzio}, \citenamefont {Devoret}, \citenamefont
  {Jiang}, \citenamefont {Girvin},\ and\ \citenamefont
  {Schoelkopf}}]{Blumoff2016}%
  \BibitemOpen
  \bibfield  {author} {\bibinfo {author} {\bibfnamefont {J.~Z.}\ \bibnamefont
  {Blumoff}}, \bibinfo {author} {\bibfnamefont {K.}~\bibnamefont {Chou}},
  \bibinfo {author} {\bibfnamefont {C.}~\bibnamefont {Shen}}, \bibinfo {author}
  {\bibfnamefont {M.}~\bibnamefont {Reagor}}, \bibinfo {author} {\bibfnamefont
  {C.}~\bibnamefont {Axline}}, \bibinfo {author} {\bibfnamefont {R.~T.}\
  \bibnamefont {Brierley}}, \bibinfo {author} {\bibfnamefont {M.~P.}\
  \bibnamefont {Silveri}}, \bibinfo {author} {\bibfnamefont {C.}~\bibnamefont
  {Wang}}, \bibinfo {author} {\bibfnamefont {B.}~\bibnamefont {Vlastakis}},
  \bibinfo {author} {\bibfnamefont {S.~E.}\ \bibnamefont {Nigg}}, \bibinfo
  {author} {\bibfnamefont {L.}~\bibnamefont {Frunzio}}, \bibinfo {author}
  {\bibfnamefont {M.~H.}\ \bibnamefont {Devoret}}, \bibinfo {author}
  {\bibfnamefont {L.}~\bibnamefont {Jiang}}, \bibinfo {author} {\bibfnamefont
  {S.~M.}\ \bibnamefont {Girvin}}, \ and\ \bibinfo {author} {\bibfnamefont
  {R.~J.}\ \bibnamefont {Schoelkopf}},\ }\href@noop {} {\bibfield  {journal}
  {\bibinfo  {journal} {Phys. Rev. X}\ }\textbf {\bibinfo {volume} {6}},\
  \bibinfo {pages} {31041} (\bibinfo {year} {2016})}\BibitemShut {NoStop}%
\end{thebibliography}%

\clearpage
\onecolumngrid
\section{Supplementary Methods}

\renewcommand{\thefigure}{S\arabic{figure}}
\renewcommand{\thetable}{S\arabic{table}}

\setcounter{figure}{0}
\setcounter{table}{0}

\subsection{Non-driven Hamiltonian and characterization of the system}

\subsubsection{System Hamiltonian}

The main system comprises 3 bosonic modes that are coupled: the qubit being measured (annihilation operator $\bq$), the readout cavity ($\bc$) and the filter mode ($\bff$). Here, we assume that the bare modes, that are capacitively coupled through a Jaynes-Cummings type of Hamiltonian, have already been diagonalized and we consider only the interaction through the Josephson Hamiltonian. Our Hamiltonian is then:

\begin{align*}
\bH &= \bH_0 + \bH_J \\
\bH_0/\hbar &= \omega_q \bq^\dag\bq + \omega_c \bc^\dag \bc + \omega_f \bff^\dag\bff \\
\bH_J/\hbar &= -\frac{E_J}{\hbar}\tilde{\cos}\left(\phi_q(\bq + \bq^\dag) + 
								\phi_c(\bc + \bc^\dag) + 
								\phi_f(\bff + \bff^\dag)\right),
\end{align*}
where $\tilde{\cos}(\phi) = \cos(\phi) + \phi^2/2$ is the cosine function without its quadratic term. This quadratic term has already been used in order to make make the transmon an harmonic oscillator (to first approximation, before expanding $\tilde{\cos}$). In the rest of the supplement, we will take the convention $\hbar = 1$. Thus, all the parameters of a Hamiltonian are angular frequencies. From this, we expand $\bH_J$ to the fourth order and keep only the non-rotating terms (which have as many $\dag$ as non-$\dag$) and we use normal ordering:

\begin{align*}
\bH &\approx \bH_0 + \bH_{\text{disp}} - \frac{\alpha}{2}\bq^{\dag 2}\bq^2 \\
\bH_{\text{disp}} &= -\chi_{qc}(\bq^\dag\bq)(\bc^\dag\bc) 
							-\chi_{qf}(\bq^\dag\bq)(\bff^\dag\bff).
\end{align*}
Here, we have considered that the cavity and the filter are entirely harmonic and that they are not coupled. $\alpha$ is the anharmonicity of the transmon qubit and allows us to later treat it as a two-level system. We link these parameters to the Josephson Hamiltonian with 

\begin{align*}
\alpha &= 2\frac{E_J}{4!}{4\choose 2}\phi_q^4 = \frac{E_J}{2}\phi_q^4 \\
\chi_{ij} &= \frac{E_J}{4!}{4\choose 1}{3\choose 1}{2\choose 1}\phi_i^2\phi_j^2 = 
			E_J\phi_i^2 \phi_j^2.
\end{align*}

\subsubsection{Characterization of the system}

In this subsection we give the characteristics of the system used for the main paper. All the data of the main paper were gathered during the same cooldown but we will also comment on observations made during other cooldowns. We summarize in table \ref{table:freqs} the frequencies of the modes and the coherence times that we measured. Then, we give the parameters of the Hamiltonian in table \ref{table:Hamiltonian}. 

\begin{table}
\centering
\caption{Frequencies and coherence times of each mode of the experimental design.}
\medskip
\begin{tabular}{ccccc}
\hline
Mode & Frequency (\SI{}{\giga \hertz}) & $T_1$ (\SI{}{\micro \second}) & $T_2$ (\SI{}{\micro \second})\\
\hline
Target Qubit & 4.982 & 90 & 30\\
Readout & 7.995 & 0.1 & - \\
Filter & 6.339  & 19 & - \\
Unread Qubit 1 & 4.686 & 140 & 30 \\
Unread Qubit 2 & 4.728 & 130 & 30 \\
\hline
\label{table:freqs}
\end{tabular}
\end{table} 

Let us comment on the results for the various qubits in table \ref{table:freqs}. First, their $T_1$'s are consistently high and are roughly constant over different cooldowns (within \SI{10}{\micro \second}). Their $T_2$'s are all the same with small variations \textit{during} each cooldowns but have larger variations \textit{between} cooldowns (between \SI{30}{\micro \second} and \SI{70}{\micro \second} over 3 cooldowns). We do not believe that the $T_2$'s were limited by photon shot-noise since we were working in the regime $\chi \ll \kappa$. We measured two other qubits in other cooldowns that we did not include here as they were not measured when we figured out the whole setup and managed to make the experiment fully work. The first extra qubit had $T_1$ of \SI{160}{\micro \second} and a $T_2$ of \SI{170}{\micro \second}. Unfortunately it ceased to work after warming up to room temperature. The second one had a $T_1$ of \SI{190}{\micro \second} and a $T_2$ of \SI{40}{\micro \second} (during a cooldown where the other qubits had a similar $T_2$). This qubit continued to work but was not measured during the last cooldown in order to free an input line.

We can now comment on the frequencies of the modes. We chose a readout cavity with a resonance frequency at \SI{8}{\giga \hertz} but this frequency only needs to be very different from the frequencies of the qubits and of the filters in order to filter the pump drive (see section on experimental setup and optical table). The qubits are designed to have a frequency around \SI{5}{\giga \hertz}, in order to be in the transmon regime ($E_J/\alpha > 50$ with $f = \sqrt{8 E_J \alpha}$) with an anharmonicity $\alpha/2\pi \approx $ \SI{250}{\mega \hertz}. The frequency of the filter mode is designed to be in between the frequencies of the qubits and of the cavity. The frequency of the filter needs to be low enough to be out of the band of a commercial band-pass filter centered at \SI{8}{\giga \hertz} but cannot be too far because the parametric coupling goes as $1/(\omega_c - \omega_f)$ (see theory section which includes the filter). This way, we can remove a lot of attenuation on the pump line and replace it by a band-pass filter centered around the frequency of the readout. This contraption allows high-power on the drive line, at the frequency of the readout mode, without having a large number of thermal excitations at the frequency of the filter mode and at the frequency of the qubit. 
\begin{table}
\centering
\caption{Coupling parameters of the Hamiltonian.}
\medskip
\begin{tabular}{ccccc}
\hline
$\alpha / 2\pi$ (\SI{}{\mega \hertz}) & $\chi_{qc}$ (\SI{}{\mega \hertz}) & $\chi_{qf}$ (\SI{}{\mega \hertz}) \\
\hline
221 & $\approx 0.1$ & 2.5\\
\hline
\label{table:Hamiltonian}
\end{tabular}
\end{table} 

In table \ref{table:Hamiltonian}, we give the coupling parameters. We see that the target qubit has a small coupling to the low-Q readout resonator and strong coupling to the intermediate-Q filter mode. Thus, the qubits are never limited by the Purcell effect and can still be strongly driven off resonance. As $\chi_{qc}$ is approximately 16 times smaller than the linewidth of the readout cavity, the value of the dispersive coupling could not be measured directly with good accuracy. A two-tone spectroscopy experiment gave $\chi_{qc} =$ \SI{70}{\kilo \hertz} but we only report the order of magnitude \SI{100}{\kilo \hertz}. Similar values were found for the unread qubits.

\subsection{Wiring diagram for strong pumping and stable cancellation}

This section comments the wiring diagram depicted on Fig. ~\ref{fig:Optical_table}.

\subsubsection{Phase and amplitude stability}

The experiment requires four phases to be stable at all times. First, the drive and the cancellation need to be tuned at the same amplitude with opposite phases. Second, the phase of the phase-sensitive amplifier (Snail-Parametric-Amplifier, SPA) \cite{Frattini2018} needs to be set such that the correct quadrature is amplified. Finally, the demodulation needs to keep the same phase over time. 

To achieve this, a single generator is set at the cavity frequency, plus \SI{50}{\mega \hertz} of sideband, and is connected to the LO ports of 3 IQ mixers and 1 mixer. The IQ mixers of the drive and the cancellation receive pulses from the DAC of an FPGA, modulated at \SI{50}{\mega \hertz} to be resonant with the frequency of the cavity. The IQ mixer of the SPA receives pulses modulated at \SI{100}{\mega \hertz} to be used as a phase-sensitive amplifier (its pump is exactly at twice the frequency of the cavity). The IQ mixers are stable both in phase and amplitude over the course of a couple of days. For safety, they are automatically re-calibrated everyday using a diagnostic port connected to a spectrum analyzer.

The signal from the cavity is amplified and demodulated using a fourth mixer. It is down-converted to \SI{50}{\mega \hertz} and digitized using the ADC of the FPGA.

\subsubsection{Strong drive that does not dephase the qubit}

The fridge was wired uniquely in order to bring a large amount of power to the base plate of the dilution refrigerator without harming the transmon qubit. For this, we use cold filters and a directional coupler. The band-pass filters are centered around the frequency of the cavity, \SI{8}{\giga \hertz}, and have a bandwidth of \SI{200}{\mega \hertz}. At room temperature we measured an attenuation of 30 dB at \SI{6.4}{\giga \hertz}, the frequency of the filter mode. The directional coupler used at 20 mK has 10 dB of coupling. The directional coupler thus sends 90\% of the power of the drive to the 4 K plate. The coupler is also used to combine the qubit drive with the strong readout drive. This contraption can be avoided by using a non-dissipative attenuator for the readout drive and a designated qubit port.

\subsubsection{Driving the unmeasured qubits}

Two other qubits are coupled to the same readout resonator. The actual aluminum cavity was obtained from a previous experiment and a picture of it is available in \cite{Blumoff2016}. In order to characterize the two unmeasured qubits, we have two more input lines available, identical to the main qubit line, without the directional coupler. During a cooldown, only one of the qubits is connected to the strong drive. 

\begin{figure}
\includegraphics{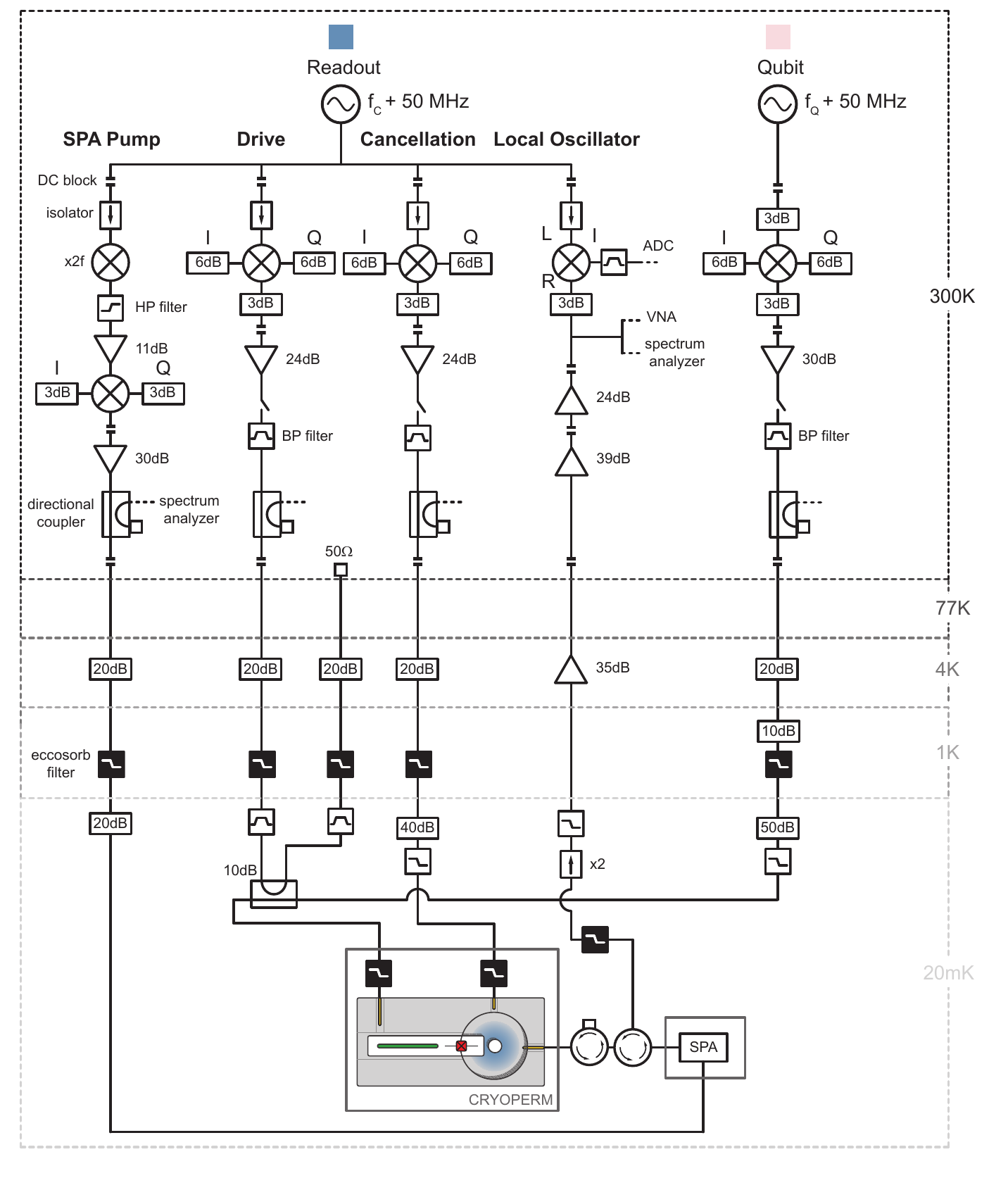}
\caption{\label{fig:Optical_table} 
Wiring Diagram
}
\end{figure}

\subsection{Characterization of the SPA}

The SPA is tuned first in phase-preserving mode, by detuning the SPA pump by \SI{3}{\mega \hertz} from twice the frequency of the cavity. The SPA has a gain of 24 dB. We characterize the bandwidth and the noise-visibility-ratio (NVR) using the spectral analyzer (see Fig.~\ref{fig:Optical_table}). We find a bandwidth of \SI{12}{\mega \hertz} and an NVR of 9 dB. The SPA pump frequency is then set at twice the frequency of the cavity to be used as a phase-sensitive amplifier. The phase of the pump is varied using the I and Q control of the IQ mixer. It is set in order to maximize the separation of the two distributions corresponding to the qubit being in $\ket{g}$ and $\ket{e}$.

\subsection{Tuning the cancellation pulse}

\begin{figure}
\includegraphics{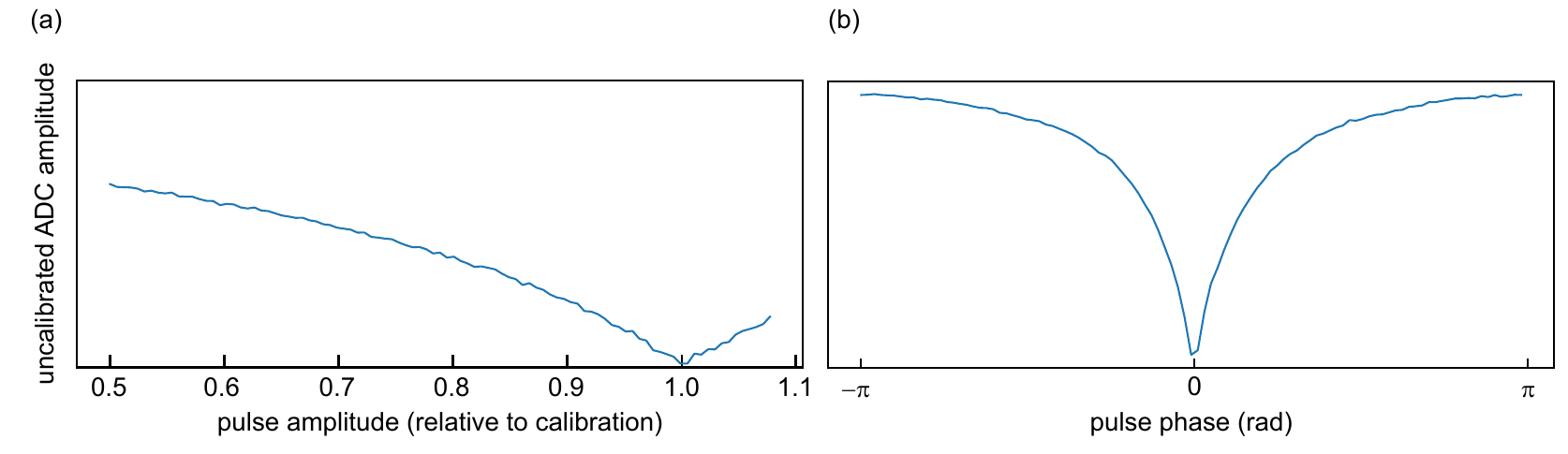}
\caption{\label{fig:cancellation} 
Tuning of the cancellation pulse. We plot the evolution of the output signal $|\bar{I}_{\ket{g}} + \bar{I}_{\ket{e}}|$ as a function of the pulse amplitude (a) and the pulse phase (b) with the same, non-calibrated, y-axis. When the amplitude of the drive and of the cancellation are the same (1.0 in (a)) and their phases are opposite (0.0 in (b)), the number of photons in the cavity is at its minimum.
}
\end{figure}

We want to minimize the number of photons present in the cavity at all times and therefore, we want the measurement histogram to be centered. For this, we calibrate the amplitude and the phase of a pulse, sent through the cancellation port, that would cancel any leakage between the drive port and the cavity. We prepare the qubit successively in $\ket{g}$ and $\ket{e}$, and we send both the cancellation drive and the readout drive. We then extract the integrated signals $\bar{I}_{\ket{g}}$ and $\bar{I}_{\ket{e}}$. On Fig.~\ref{fig:cancellation} we plot $|\bar{I}_{\ket{g}} + \bar{I}_{\ket{e}}|$ as a function of the pulse amplitude and phase (the y-axis is not calibrated). When the cancellation pulse has the same amplitude as the drive but opposite phase, this quantity is at a minimum. This minimum is stable over the course of a couple of days. The histogram presented in Fig. 2 of the main paper shows that the corresponding distributions are indeed centered around the origin. In practice, this protocol is automatically repeated multiple times until the minimum is below a given threshold. 

\subsection{Parametric pumping of the Josephson Hamiltonian}

In this section, we discuss the theory behind the effective Hamiltonian given in the main text. In a first part we give the theory without considering the filter mode, which corresponds to the main text. Then, we talk about how to take into account the filter mode, as it was designed in the experiment.

\subsubsection{Without the filter mode}

The non-driven Hamiltonian of such a system was given in the first section. Here, we first ignore the filter mode $\bff$. The total Hamiltonian is then 

\begin{align*}
\bH = \omega_q \bq^\dag\bq + \omega_c \bc^\dag\bc - E_J\tilde{\cos}\left(
\phi_q(\bq+\bq^\dag) + \phi_c(\bc + \bc^\dag)\right) + \epsilon(t)\bq^\dag + \epsilon(t)^*\bq,
\end{align*}
where $\epsilon(t)$ represents the amplitude of the drive. It is time dependent and modulated at the frequency of the cavity. We write it $\epsilon(t) = \bar{\epsilon}(t)e^{i\omega_c t}$, where $\bar{\epsilon}(t)$ is a slowly varying envelope. Sometimes, for clarity, we will use $\bm{\phi} = \phi_q(\bq+\bq^\dag) + \phi_c(\bc + \bc^\dag)$. 

The Langevin equation for the qubit is 

\begin{align*}
\dot{\bq} = - i \omega_q\bq + iE_J[\bq, \tilde{\cos}(\bm\phi)] - i\epsilon(t) - 
\frac{\Gamma}{2}\bq,
\end{align*}
where $\Gamma$ is the decay rate of the qubit. 

We perform a change of frame for the qubit and we define the operator $\bar{\bq} = (\bq + \xi(t))e^{i\omega_q t}$, where $\xi(t)$ is, for now, an arbitrary displacement that depends on time. This transformation is unitary because $[\bar{\bq}, \bar{\bq}^\dag] = 1$. We define the new phase operator $\bar{\bm\phi} = \phi_q(\bar\bq e^{-i\omega_q t} + \bar\bq^\dag e^{i\omega_q t} - \xi(t) - \xi(t)^*) + \phi_c(\bc + \bc^\dag)$, where we have applied the inverse transform on $\bq$. When the cosine term is expanded in $\bar{\bm\phi}$, it gives new \textit{3rd order} terms that are time dependent and can be made resonant, depending on the modulation of $\xi$. The Langevin equation for $\bar\bq$ is 

\begin{align*}
\dot{\bar\bq} &= (\dot{\bq} + i\omega_q \bq
 + \dot{\xi}(t) + i \omega_q \xi(t))e^{i \omega_q t} \\
 &= (-i\omega_q \bq + i \omega_q \bq + iE_J [\bq, \tilde{\cos}(\bar{\bm\phi})]
 - \Gamma/2\bq - i\epsilon(t) +\dot\xi(t) + i\omega_q\xi(t))e^{i\omega_q t} \\
 &= iE_J[\bar\bq, \tilde\cos(\bar{\bm\phi})] - \Gamma/2\bar\bq + (\dot\xi(t) + 
 i\omega_q \xi(t) + \Gamma/2\xi(t) - i\epsilon(t))e^{i\omega_q t},
\end{align*}
In order to focus on the dynamic induced by the pumped Josephson Hamiltonian, we \textit{choose} the displaced frame $\xi(t)$ such that it solves the differential equation

\begin{align*}
\dot\xi(t) = -(\Gamma/2 + i\omega_q) \xi(t) + i\epsilon(t).
\end{align*} 

As an illustrative example, we study the case $\bar\epsilon(t) = \bar\epsilon$ (constant pulse at frequency $\omega_c$). In this case, 

\begin{align*}
\xi(t) &= \frac{i\bar\epsilon}{\Gamma/2 - i\Delta}e^{i\omega_c t} + Ae^{-(\Gamma/2 + i\omega_q) t} \\
&\approx -\frac{\bar\epsilon}{\Delta}e^{i\omega_c t} + Ae^{-(\Gamma/2 + i\omega_q) t},
\end{align*}
where $\Delta = \omega_c - \omega_q$. This displaced frame contains an off-resonant part, at frequency $\omega_c$, which will lead to the desired longitudinal coupling. The second part is resonant, at $\omega_q$ and decays slowly with rate $\Gamma$, which is the energy relaxation of the qubit. This term is undesirable and causes unwanted transitions during the parametric pumping process. It is due to the fact that the pulse is instantaneous. It is analogous to Landau-Zenner transitions when a parameter of the Hamiltonian is varied in a non-adiabatic way. Fortunately, it can easily be suppressed when the ring-up time of the envelope $\bar\epsilon(t)$ is slow compared to $1/\Delta$ (which is \SI{3}{\giga \hertz} in our experiment). The displaced frame is thus 

\begin{align*}
\xi(t) = -\frac{\bar{\epsilon}}{\Delta}e^{i\omega_c t}.
\end{align*}

We now go to the rotating frame of the cavity, such that the only remaining part is the driven Josephson Hamiltonian

\begin{align*}
\bar\bH_J = -E_J\tilde\cos(\phi_q(\bar\bq e^{-i\omega_q t} + \bar\bq^\dag e^{i\omega_q t} - \xi(t) - \xi(t)^*) + \phi_c(\bc e^{-i\omega_c t} + \bc^\dag e^{i\omega_c t})).
\end{align*}

As usual, we expand the Josephson Hamiltonian to the fourth order and keep only the non-rotating terms. The main ones are the usual Kerr and dispersive coupling terms, whose expressions are given in the first section. The most relevant resonant term created by the drive is a resonant longitudinal coupling. Up to a displacement to center the conditional displacement, we obtain

\begin{align*}
\bH_{\text{eff}} = -\frac{\alpha}{2}\bq^{\dag 2}\bq^2 - \chi(\bq^\dag\bq)(\bc^\dag\bc) + \zeta(t)(\bq^\dag\bq-\textbf{1}/2)(\bc + \bc^\dag),
\end{align*}
where we find

\begin{align*}
|\zeta| &= {4\choose 1}{3\choose 1}{2\choose 1}{1\choose 1}\frac{E_J}{4!} \phi_q^3\phi_c|\xi| \\
&= \sqrt{2\alpha\chi}\left\lvert\frac{\bar\epsilon}{\Delta}\right\rvert.
\end{align*}

Notice that we will drop the bars on $\bq$ and $\bc$ for the rest of the supplement. In practice we observe that the qubit starts heating when $\xi \approx 1$, which makes the measurement non-QND. However, for $\xi = 1$ we already have $|\zeta| > \kappa$ and therefore we can still make a fast measurement.

\subsubsection{With the filter mode}

The role of the filter mode is explicit in the main text. We can easily take it into account in the previous derivation. We do not reproduce all of it, as it is extremely similar. The drive is now replaced by $\epsilon(t)\bff^\dag + \epsilon(t)\bff$, and the role of $\Delta$ is now $\Delta_f = \omega_c - \omega_f$. The same derivation leads to 

\begin{align*}
|\zeta| = \sqrt{\chi_{qf}\chi_{qc}}\left\lvert\frac{\bar\epsilon}{\Delta_f}\right\rvert.
\end{align*}

We see that we need to choose $\chi_{qf}$ large in order to get $\sqrt{\chi_{qf}\chi_{qc}}$ of order $\kappa$. However, we need to keep $\chi_{qf}$ small enough such that the dissipation of the filter (due to its coupling to the drive pin) does not limit the lifetime of the qubit through the Purcell effect. There is a large margin of optimization in these parameters for future implementations of resonant longitudinal couplings. The single-mode Purcell limit $T_P$ on the lifetime of the qubit due to the filter mode is $T_P = T_f \alpha/\chi_{qf}$, where $T_f$ is the lifetime of the filter mode. Using the parameters from tables \ref{table:freqs} and \ref{table:Hamiltonian}, we get $T_P > $ \SI{1}{\milli \second}. This shows that we could use a larger value of $\chi_{qf}$ and thus have a bigger longitudinal coupling for the same drive strength.

\subsection{Langevin equation}

We derive the dynamic of the cavity under the dispersive and the resonant longitudinal couplings. In the following, we use $\boldsymbol{\sigma}_Z = 2(\bq^\dag\bq - \boldsymbol{1}/2)$, the usual Pauli operator, and we drop the Kerr term $\bq^{\dag 2}\bq^2$.

\subsubsection{Dispersive}

The Langevin equation for the dispersive readout in reflection is 

\begin{align}
\dot{\ba} = -i\frac{\chi}{2}\sigma_Z\ba - \frac{\kappa}{2}\ba + \sqrt{\kappa}\ba_{in},
\end{align}
with the input-output relation $\ba_{out} = \ba_{in} + \sqrt{\kappa}\ba$. In our case, $\langle \ba_{in} \rangle = -\epsilon / \sqrt{\kappa}$ with $\epsilon$ chosen real. We solve for $t$ and get two semi-classical solutions that are coherent states with amplitude

\begin{align*}
\alpha(t, \sigma_Z) = \frac{2\epsilon}{\kappa(1+i\chi\sigma_Z/\kappa)}\left(1 - e^{-\kappa(1+i\chi\sigma_Z/\kappa)t/2}\right).
\end{align*}

We note $\phi_{qb} = \arctan{\chi/\kappa}$. Hence, 

\begin{align*}
\alpha(t, \sigma_Z) = \frac{2\epsilon}{\kappa}\cos{(\phi_{qb})}e^{-i\phi_{qb}\sigma_Z}\left(1 - e^{-\kappa(1+i\chi\sigma_Z/\kappa)t/2}\right).
\end{align*}

Using the input-output relation we have the output field coherent state

\begin{align}
\alpha_{out}(t, \sigma_Z) &= \frac{2\epsilon}{\sqrt{\kappa}}\left(\cos(\phi_{qb})e^{-i\phi_{qb}\sigma_Z}\left(1 - e^{-\kappa(1+i\chi\sigma_Z/\kappa)t/2}\right) - \frac{1}{2}\right) \\
&= \frac{\epsilon}{\sqrt{\kappa}}e^{-2i\phi_{qb}}\left(1 - 2\cos(\phi_qb)e^{-\kappa(1+i\chi\sigma_Z/\kappa)t/2 + i\phi_{qb}\sigma_Z}\right).
\end{align}

\subsubsection{Longitudinal}

The Hamiltonian we consider is $\bH/\hbar = \zeta(t)/2\boldsymbol{\sigma_Z}(\ba + \ba^\dag)$ with $
\zeta$ real and constant (this is a choice here). The Langevin equation is 

\begin{align}
\dot{\ba} = -i\frac{\zeta}{2}\sigma_Z - \frac{\kappa}{2}\ba.
\end{align}

The corresponding semi-classical solutions are two coherent states with amplitude

\begin{align}
\alpha(t, \sigma_Z) = i\sigma_Z\frac{\zeta}{\kappa}\left(1 - e^{-\kappa t /2}\right).
\label{eq:alpha_Z}
\end{align}

Since the signal is obtained by modulating the coupling, we have $\langle\ba_{in}\rangle = 0$ so $\alpha_{out}(t, \sigma_Z) = \sqrt{\kappa}\alpha(t, \sigma_Z)$.

\subsection{Signal to Noise Ratio}

\subsubsection{General notations}

The measurement operator for a homodyne detection is $\bM(\tau) = \sqrt{\kappa}\bigints_0^\tau {\left[\ba^\dag_{out}e^{i\varphi} + \ba_{out}e^{-i\varphi}\right]K(t)dt}$, with $\varphi$ the demodulation angle, fixed by the pump of the phase-sensitive amplifier. $K(t)$ is a time-dependent demodulation envelope. For example, for a Boxcar demodulation we have $K(t) = 1$ at any time. The corresponding noise operator is $\bM_{Ni} = \bM - \langle\bM\rangle_i$, where the index $i = g, e$ is for the qubit in its ground or excited state and $\langle \bM \rangle_i = \bra{\alpha_{out}(t, i)} \bM \ket{\alpha_{out}(t, i)}$. We define the signal-to-noise ratio as 

\begin{align}
\label{eq:SNR}
\text{SNR} = \sqrt{\frac{|\langle\bM\rangle_e - \langle\bM\rangle_g|^2}{\langle\bM_{Ne}^2\rangle_e + \langle\bM_{Ng}^2\rangle_g}}
\end{align} 

Notice that the noise is simply due to the commutation relation $\left[\ba_{out}(t), \ba_{out}^\dag(t')\right] = \delta(t, t')$. For example, for a demodulation using a constant filter, we get $\langle \bM_{N i}^2 \rangle = \kappa \tau$ and the numerator is the only part that depends on what coupling is used to read the qubit.

It has been shown \cite{Gambetta2007, Ryan2015, Bultink2018} that the SNR is optimal when $K(t) = \langle \ba_{out, e} - \ba_{out, g}\rangle^*$, and that in that case, the SNR is given by 

\begin{align}
\text{SNR} = \sqrt{2\int_0^\tau|\alpha_{out, e} - \alpha_{out, g}|^2dt}.
\end{align}

\subsubsection{Dispersive (Boxcar)}

Looking at the numerator of eq. \ref{eq:SNR}, we have $\langle\bM\rangle_e - \langle\bM\rangle_g = \sqrt{\kappa}\bigints_0^\tau 
\left[\left(\alpha_{out, e}^* - \alpha_{out, g}^*\right)e^{i\varphi} + \left(
\alpha_{out, e} - \alpha_{out, g}\right) e^{-i\varphi}\right]dt$. In the case of the dispersive readout, 

\begin{align}
\alpha_{out, e} - \alpha_{out, g} = \frac{\epsilon}{\sqrt{\kappa}}\left(2i\sin(2\phi_{qb}) - 4i\cos(\phi_{qb})\sin(\phi_{qb}+\chi t /2)e^{-\kappa t /2}\right),
\label{eq:disp_dist}
\end{align}
and so $\alpha_{out, e}^* - \alpha_{out, g}^* = -( \alpha_{out, e} - \alpha_{out, g})$. Then,

\begin{align*}
\langle\bM\rangle_e - \langle\bM\rangle_g &= 4\epsilon\sin(\varphi)\sin(2\phi_{qb})
\bigints_0^\tau\left[ 1 - 2\cos(\phi_{qb})\frac{\sin(\phi_{qb} + \chi t/2)}{\sin(2\phi_{qb})}e^{-\kappa t/2}\right]dt \\
&= 4\epsilon\sin(\varphi)\sin(2\phi_{qb})\left[\tau - \frac{4\cos^2(\phi_{qb})}{\kappa}\left(1 - \frac{\sin(2\phi_{qb}+\frac{\chi\tau}{2})}{\sin(2\phi_{qb})}e^{-
\kappa\tau/2}\right)\right].
\end{align*}

Together with the noise operator defined above and taking the expression of the SNR we find 

\begin{align}
\text{SNR} =  \sqrt{8}\frac{\epsilon}{\kappa}\sin(\varphi)\sin(2\phi_{qb})\sqrt{\kappa\tau}\left[1 - \frac{4\cos^2(\phi_{qb})}{\kappa\tau}\left(1 - \frac{\sin(2\phi_{qb}+\frac{\chi\tau}{2})}{\sin(2\phi_{qb})}e^{-
\kappa\tau/2}\right)\right].
\end{align}

This corresponds to the derivation done in \cite{Didier2015}. For the case $\chi = \kappa$,

\begin{align}
\text{SNR} = \sqrt{8}\frac{\epsilon}{\kappa}\sqrt{\kappa \tau}\left(
1 - \frac{2}{\kappa\tau}\left(1 - \cos(\frac{1}{2}\kappa\tau)e^{-\kappa\tau/2}\right)\right).
\end{align}

From this, we have two regimes. When $\kappa\tau\gg1$ we get $\text{SNR}\propto\sqrt{\kappa\tau}$ which is the best steady-state regime possible (without squeezing). When $\kappa\tau\ll1$ we have $\text{SNR}\propto(\kappa\tau)^{5/2}$. We will demonstrate that this is slower than for the conditional displacement obtained with the longitudinal coupling.

\subsubsection{Dispersive (Optimal)}

We start from eq. \ref{eq:disp_dist} where we notice that all the information is along one quadrature. We find that

\begin{align}
|\alpha_{out,e} - \alpha_{out, g}|^2 &= \frac{4\epsilon^2}{\kappa}\left(\sin^2(2\phi_{qb}) - 4\cos(\phi_{qb})\sin(2\phi_{qb})e^{-\kappa t/2}\sin(\phi_{qb} + \frac{\chi t}{2}) + 2e^{-\kappa t}\cos^2(\phi_{qb})(1 - \cos(2\phi_{qb} + \chi t))\right).
\end{align}

We denote the two integrals $I_1$ and $I_2$:

\begin{align}
I_1 &= \int_0^\tau\sin(\phi_{qb} + \frac{\chi t}{2})e^{-\kappa t /2}dt \\
&= \frac{2}{\kappa}\cos(\phi_{qb})\sin(2\phi_{qb})\left(1 - e^{-\frac{\kappa\tau}{2}}\frac{\sin(2\phi_{qb} + \frac{\chi\tau}{2})}{\sin(2\phi_{qb})}\right) \\
I_2 &= \int_0^\tau \cos(2\phi_{qb} + \chi t)e^{-\kappa t}dt\\
&= \frac{\cos(\phi_{qb})}{\kappa}\left(\cos(3\phi_{qb}) - e^{-\kappa \tau}\cos(3\phi_{qb} + \chi\tau)\right)
\end{align}

Finally we get 

\begin{align}
\text{SNR} = \sqrt{8}\frac{\epsilon}{\sqrt{\kappa}}\left(\sin^2(2\phi_{qb})\tau - 4\cos(\phi_{qb})\sin(2\phi_{qb})I_1 + \frac{2}{\kappa}\cos^2(\phi_{qb})(1 - e^{-\kappa\tau})-2\cos^2(\phi_{qb})I_2 \right)^{1/2}
\end{align}

We use this expression with $\chi = \kappa$ on Fig. 2a of the main paper. Similarly to the non-optimal case, when $\kappa \tau \gg 1$ we have $\text{SNR} \propto \sqrt{\kappa\tau}$ (the envelope does not make it sub-optimal). 

\subsubsection{Longitudinal (Boxcar)}

Using eq. \ref{eq:alpha_Z} we find $\alpha_{out, e} - \alpha_{out, g} = \frac{2\zeta}{\sqrt{\kappa}}\left(1 - e^{-\kappa t /2}\right) = - (\alpha_{out, e} - \alpha_{out, g})^*$, which leads to:

\begin{align}
\text{SNR} = \sqrt{8}\frac{\zeta}{\kappa}\sqrt{\kappa \tau}\left(1 - \frac{2}{\kappa\tau}\left(1-e^{-\kappa\tau/2}\right)\right).
\end{align}

Here, there are also two different regimes. First, when $\kappa\tau\gg 1$, $\text{SNR}\propto\sqrt{\kappa\tau}$. The conditional displacement readout is also optimal in the sense that the measurement rate is equal to the dephasing rate for an efficiency 1 (see section on dephasing rate and efficiency). Second, for $\kappa\tau \ll 1$ we have $\text{SNR}\propto \left(\kappa\tau\right)^{3/2}$, which is faster than the dispersive readout when $\chi = \kappa$. Our calculation shows that it is, however, similar if $\chi \ll \kappa$ and that the number of photons is adequately changed. Nevertheless, adapting the number of photons is not convenient to readout multiple qubits with the same readout resonator.

\subsubsection{Longitudinal (Optimal)} 

This expression for the SNR is the one we use to fit to our data on Fig. 2a of the main paper. We find 

\begin{align}
\text{SNR} &= \sqrt{8}\frac{\zeta}{\kappa}\left(\kappa\tau - 4\left(1-e^{-\kappa\tau/2}\right) + \left(1-e^{-\kappa\tau}\right)\right)^{1/2}.
\end{align}
With this expression, we also find $\text{SNR}(\kappa\tau\ll 1)\propto(\kappa\tau)^{3/2}$ and $\text{SNR}(\kappa\tau\gg 1)\propto\sqrt{\kappa\tau}$.

\subsection{Dephasing rate and efficiency}

\subsubsection{Theory to determine the efficiency $\eta$}

In this section, we want to demonstrate that we can always relate the dephasing rate of the qubit to the SNR (with optimal envelope) and thus we can always find the efficiency in a similar way as shown in \cite{Bultink2018}. We show that this strategy does not depend on what Hamiltonian is used to read the qubit. However, we make the assumption that the state of the system (qubit + cavity) is at all times in a state

\begin{align}
\rho = a\ket{g, \alpha_g(t)}\bra{g, \alpha_g(t)} + b\ket{e, \alpha_e(t)}\bra{e, \alpha_e(t)} +
c\ket{e, \alpha_e(t)}\bra{g, \alpha_g(t)} + d\ket{g, \alpha_g(t)}\bra{e, \alpha_e(t)}
\end{align}

From this we usually perform a polaron transform \cite{Gambetta2008, Didier2015}. Here, we first displace the cavity to the barycenter of the two coherent states $(\alpha_g + \alpha_e)/2$. In this frame, the state of the cavity is $\pm(\alpha_g - \alpha_e)/2$, depending on the state of the qubit. The state of the cavity is thus symmetric around the origin. Our "polaron" tranform is thus
	
\begin{align}
\ba \rightarrow \ba + \frac{\alpha_g(t) + \alpha_e(t)}{2} + \frac{\alpha_g(t) - \alpha_e(t)}{2}\boldsymbol{\sigma_Z}.
\end{align}

The commutation relations of $\ba$ are not changed by this transformation and so it is unitary. When we apply it to the Lindblad superoperator, it highlights the dephasing of the qubit. We have 

\begin{align}
\mathcal{D}[\ba]\rho &= 2\ba^\dag\rho\ba - \ba^\dag\ba\rho - \rho\ba^\dag\ba \\
&\rightarrow \mathcal{D}[\ba + \frac{\alpha_g(t) + \alpha_e(t)}{2} + \frac{\alpha_g(t) - \alpha_e(t)}{2}\boldsymbol{\sigma_Z}]\rho \\
&= \mathcal{D}[\ba]\rho + \frac{1}{4}|\alpha_g(t) - \alpha_e(t)|^2\mathcal{D}[\boldsymbol{\sigma_Z}]\rho + ...,
\end{align}

where the last part is either unitary or neglected using the rotating-wave approximation. We make the identification 

\begin{align}
\frac{\kappa}{8}|\alpha_g(t) - \alpha_e(t)|^2\mathcal{D}[\boldsymbol{\sigma_Z}]\rho = \frac{1}{4}\Gamma_m\mathcal{D}[\boldsymbol{\sigma_Z}]\rho,
\end{align}

which gives the same relation as in \cite{Bultink2018}, generalized to an arbitrary signal. We define the total dephasing $\gamma_m$ such that the coherence of the density matrix of the qubit is given by $|\rho_{ge}(\tau)| = e^{-\gamma_m}|\rho_{ge}(0)|$. We get
\begin{align}
\gamma_m &= \frac{\kappa}{2}\int_0^\tau|\alpha_g(t) - \alpha_e(t)|^2dt \\
&= \frac{1}{2}\int_0^\tau|\alpha_{out, g} - \alpha_{out, e}|^2dt \\
&= \frac{1}{4}\text{SNR}^2.
\end{align}

We define the efficiency as the part of the qubit dephasing that does not get captured by the SNR. We define the experimental SNR as $\sqrt{2\eta\int_0^\tau|\alpha_{out, g} - \alpha_{out, e}|^2dt}$. Now, we can determine the efficiency using

\begin{align}
\eta = \frac{\text{SNR}^2}{4\gamma_m}.
\end{align}

This formula is similar to what has been found before \cite{Bultink2018} but here it can be used for our conditional displacement readout.

\subsubsection{Experimental determination of $\eta$}

\begin{figure}
\includegraphics{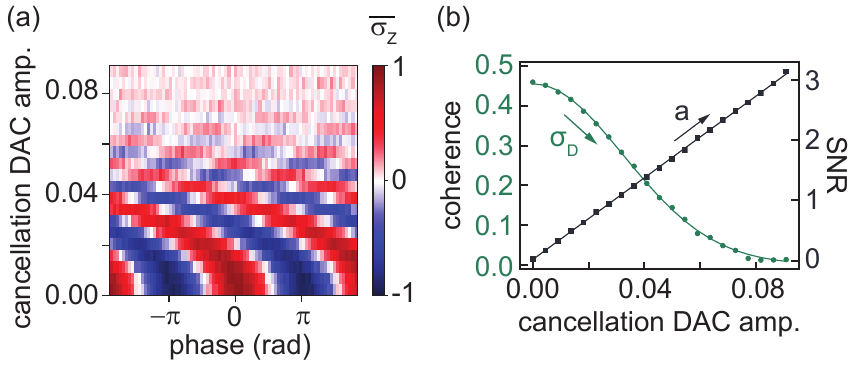}
\caption{\label{fig:Efficiency} 
Calibration of the efficiency. (a) Ramsey fringes as a function of the DAC amplitude used for the cancellation pulse. Their amplitude decays with the DAC amplitude and their phase has an off-set due to the Stark-shift. (b) Coherence of the target qubit (green) and SNR (gray) as a function of the same DAC amplitude. The plain lines correspond to a Gaussian fit for the coherence and a linear fit for the SNR. $\sigma_\text{D}$ is the standard deviation of the Gaussian fit and a is the slope of the linear fit.
}
\end{figure}

We measure the efficiency using a Ramsey experiment with fixed length in which we introduce a measurement pulse of variable strength. The sequence is similar to the one depicted in Fig. 4 of the main paper, but where the $\pi/2$-pulses are on the target qubit. The measurement sequence is also slightly different from the one used in Fig. 4. We use a \SI{500}{\nano \second} pulse and \SI{1}{\micro \second} wait time and we do not use the depletion pulse (in case it is not correctly tuned for each measurement strength). The results are depicted on Fig.~\ref{fig:Efficiency}. The amplitude of the Ramsey fringes has a Gaussian decay with standard deviation $\sigma_D$ and the SNR, demodulated with the optimal envelope, grows linearly with the DAC amplitude. This is expected from the theory in the previous subsection. From the two parameters of the fits we compute the efficiency 

\begin{equation*}
\eta = \frac{\sigma_\text{D}^2\text{a}^2}{2} = 0.6.
\end{equation*}

\end{document}